\documentclass[tighten, notrackchanges, twocolumn]{aastex62}

\usepackage{amsmath}
\usepackage{amssymb}
\usepackage{amssymb}
\usepackage{mathtools}
\usepackage{mathrsfs}
\usepackage{acronym} 
\usepackage{morefloats}
\usepackage{longtable}
\usepackage{afterpage}
\usepackage{ragged2e}
\usepackage{multirow}
\usepackage[usestackEOL]{stackengine}

\hypersetup{linkcolor=red,citecolor=blue,urlcolor=magenta}

\usepackage{soul}

\usepackage[makeroom]{cancel}

\graphicspath{{figs/}}

\usepackage{xspace}

\newcommand*{\xmm}{XMM-Newton\xspace}
\newcommand*{\Multinest}{\textsc{MultiNest}\xspace}
\newcommand*{\Pymultinest}{\textsc{PyMultiNest}\xspace}

\newcommand*{\msol}{M$_\odot$\xspace}
\newcommand*{\joo}{PSR~J0030$+$0451\xspace}
\newcommand*{\jos}{PSR~J0740$+$6620\xspace}
\newcommand*{\jof}{PSR~J0437$-$4715\xspace}

\newcommand*{\CI}{CI$_{68}$\xspace}

\acrodef{X-PSI}{X-ray Pulse Simulation and Inference}
\acrodef{MSP}{millisecond pulsar}
\acrodef{PPM}{Pulse Profile Modeling}
\acrodef{EoS}{Equation of State}
\acrodef{ISS}{International Space Station}
\acrodef{NS}{neutron star}
\acrodefplural{NS}[NSs]{neutron stars}
\acrodef{NICER}{Neutron Star Interior Composition Explorer}
\acrodef{XMM}{XMM-Newton}
\acrodef{GTI}{Good Time Interval}
\acrodefplural{GTI}[GTIs]{Good Time Intervals}

\newcommand{\TT}[1]{\texttt{#1}}

\defcitealias{Riley19}{R19}
\defcitealias{Riley21}{R21}
\defcitealias{bogdanov19a}{B19}
\defcitealias{Vinciguerra23a}{V23a}
\defcitealias{Vinciguerra23b}{V23b}
\defcitealias{Salmi22}{S22}
\defcitealias{Salmi23}{S23}

\received{}
\revised{}
\accepted{}
\published{}

\shorttitle{A NICER VIEW OF \jof}
\shortauthors{Choudhury~et~al.}

\begin{document}

\title{A NICER VIEW OF THE NEAREST AND BRIGHTEST MILLISECOND PULSAR: \jof}

\correspondingauthor{D.~Choudhury}
\email{D.Choudhury@uva.nl}

\author[0000-0002-2651-5286]{Devarshi~Choudhury}
\affil{Anton Pannekoek Institute for Astronomy, University of Amsterdam, Science Park 904, 1090GE Amsterdam, the Netherlands}

\author[0000-0001-6356-125X]{Tuomo~Salmi}
\affil{Anton Pannekoek Institute for Astronomy, University of Amsterdam, Science Park 904, 1090GE Amsterdam, the Netherlands}

\author[0000-0003-3068-6974]{Serena~Vinciguerra}
\affil{Anton Pannekoek Institute for Astronomy, University of Amsterdam, Science Park 904, 1090GE Amsterdam, the Netherlands}

\author[0000-0001-9313-0493]{Thomas~E.~Riley}
\affil{Anton Pannekoek Institute for Astronomy, University of Amsterdam, Science Park 904, 1090GE Amsterdam, the Netherlands}

\author[0000-0002-0428-8430]{Yves~Kini}
\affil{Anton Pannekoek Institute for Astronomy, University of Amsterdam, Science Park 904, 1090GE Amsterdam, the Netherlands}

\author[0000-0002-1009-2354]{Anna~L.~Watts}
\affil{Anton Pannekoek Institute for Astronomy, University of Amsterdam, Science Park 904, 1090GE Amsterdam, the Netherlands}

\author[0000-0002-9407-0733]{Bas Dorsman}
\affil{Anton Pannekoek Institute for Astronomy, University of Amsterdam, Science Park 904, 1090GE Amsterdam, the Netherlands}

\author[0000-0002-9870-2742]{Slavko~Bogdanov} 
\affil{Columbia Astrophysics Laboratory, Columbia University, 550 West 120th Street, New York, NY 10027, USA}

\author[0000-0002-6449-106X]{Sebastien~Guillot}
\affil{IRAP, CNRS, 9 avenue du Colonel Roche, BP 44346, F-31028 Toulouse Cedex 4, France}
\affil{Universit\'{e} de Toulouse, CNES, UPS-OMP, F-31028 Toulouse, France.}

\author[0000-0002-5297-5278]{Paul~S.~Ray}
\affil{Space Science Division, U.S. Naval Research Laboratory, Washington, DC 20375, USA}

\author[0000-0002-2035-4688]{Daniel~J.~Reardon}
\affil{Centre for Astrophysics and Supercomputing, Swinburne University of Technology, P.O. Box 218, Hawthorn, Victoria 3122, Australia}
\affil{OzGrav: The Australian Research Council Centre of Excellence for Gravitational Wave Discovery, Hawthorn VIC 3122, Australia} 

\author[0000-0003-4815-0481]{Ronald A.~Remillard}
\affil{MIT Kavli Institute for Astrophysics \& Space Research, MIT, 70 Vassar St., Cambridge, MA 02139, USA}

\author[0000-0002-7177-6987]{Anna~V.~Bilous}
\affil{ASTRON, the Netherlands Institute for Radio Astronomy, Postbus
2, 7990 AA Dwingeloo, The Netherlands}

\author[0000-0002-1169-7486]{Daniela~Huppenkothen}
\affil{SRON Netherlands Institute for Space Research, Niels Bohrlaan 4, 2333 CA Leiden, the Netherlands}

\author[0000-0002-5907-4552]{James M.~Lattimer}
\affil{Department of Physics and Astronomy, Stony Brook University, Stony Brook, NY 11794-3800, USA}

\author[0000-0002-9626-7257]{Nathan Rutherford}
\affiliation{Department of Physics and Astronomy, University of New Hampshire, Durham, New Hampshire 03824, USA}

\author{Zaven~Arzoumanian}
\affil{X-Ray Astrophysics Laboratory, NASA Goddard Space Flight Center, Code 662, Greenbelt, MD 20771, USA}

\author[0000-0001-7115-2819]{Keith~C.~Gendreau}
\affil{X-Ray Astrophysics Laboratory, NASA Goddard Space Flight Center, Code 662, Greenbelt, MD 20771, USA}

\author[0000-0003-4357-0575]{Sharon~M.~Morsink}
\affil{Department of Physics, University of Alberta, 4-183 CCIS, Edmonton, AB, T6G 2E1, Canada}

\author[0000-0002-6089-6836]{Wynn~C.~G.~Ho}
\affil{Department of Physics and Astronomy, Haverford College, 370 Lancaster Avenue, Haverford, PA 19041, USA}

\begin{abstract}

We report Bayesian inference of the mass, radius and hot X-ray emitting region properties - using data from the Neutron Star Interior Composition ExploreR (NICER) - for the brightest rotation-powered millisecond X-ray pulsar \jof.  Our modeling is conditional on informative tight priors on mass, distance and binary inclination obtained from radio pulsar timing using the Parkes Pulsar Timing Array (PPTA) \citep{radio_prior}, and we use NICER background models to constrain the non-source background, cross-checking with data from XMM-Newton. We assume two distinct hot emitting regions, and various parameterized hot region geometries that are defined in terms of overlapping circles; while simplified, these capture many of the possibilities suggested by detailed modeling of return current heating. For the preferred model identified by our analysis we infer a mass of $M = 1.418 \pm 0.037$ \msol (largely informed by the PPTA mass prior) and an equatorial radius of $R = 11.36^{+0.95}_{-0.63}$ km, each reported as the posterior credible interval bounded by the 16\% and 84\% quantiles. This radius favors softer dense matter equations of state and is highly consistent with constraints derived from gravitational wave measurements of neutron star binary mergers. The hot regions are inferred to be non-antipodal, and hence inconsistent with a pure centered dipole magnetic field.   

\end{abstract}


\section{Introduction}\label{sec:intro}

NICER, the Neutron Star Interior Composition Explorer \citep{Gendreau2016}, collects high time resolution X-ray spectral-timing data of rotation-powered millisecond pulsars (MSPs) to build up high quality phase and spectrally resolved pulse profiles \citep{bogdanov19a}.  Relativistic effects due to the neutron star's rapid spin and strong gravitational field affect the radiation emitted from the hot magnetic polar caps of the pulsar in a well-understood fashion \citep[see][and references therein]{Bogdanov19b}.  Via Pulse Profile Modeling (PPM) we can thus infer the pulsar's mass and radius as well as geometric properties such as the size and location of the emitting polar caps \citep{Watts2019,Bogdanov21}. Mass and radius measurements constrain the Equation of State (EoS) of the supranuclear dense matter in neutron star cores \citep[see e.g.][]{Lattimer07}, while the hot polar cap properties map the magnetic field structure \citep[e.g.][]{Bilous_2019}.  

To date the NICER collaboration have reported PPM results for two MSPs:  \joo \citep{Miller19,Riley19,Salmi23,Vinciguerra23b} and the heavy pulsar \jos \citep{Miller21,Riley21,Salmi22,Salmi23,Salmi24,Dittmann24}.  Radius uncertainties for both are at present $\sim \pm 10$ \% (68\% credible interval), with this expected to improve as more photons are accumulated \citep{Lo13,Psaltis14}.  \joo is the second brightest source in the NICER PPM target list, but is an isolated pulsar for which there is no prior information on the mass.  \jos is in a binary and has a well-constrained mass derived from radio pulsar timing \citep{Fonseca21} that is used as a prior in the pulse profile modeling of NICER data. With a high neutron star mass $\sim 2.1$ \msol, it is a very interesting source in terms of potential constraints on dense matter.  However, being more distant, it is fainter, which makes it more challenging to obtain sufficient photons for tight constraints on radius. NICER's results for these two sources have already started to place interesting constraints on both dense matter \citep[see, e.g.,][]{Miller21,Raaijmakers2021,Biswas2022,Annala2023,Takatsy23} and pulsar magnetic field geometry and emission mechanisms \citep[see, e.g.,][]{Bilous_2019,Chen2020,Kalapotharakos2021,Carrasco2023}.  NICER PPM analysis has been supported by a programme of testing and simulation \citep{Bogdanov19b,Bogdanov21,Vinciguerra23a,Choudhury24_raytracing}.

The rotation powered pulsar \jof, which is the focus of this Letter, was discovered by \citet{Johnston93} and is the closest observed MSP and the brightest source in the NICER PPM target list. Its brightness and proximity allows for precise measurements of astrometric and orbital parameters from radio pulsar timing, and it is possible to obtain a tightly-constrained distance value of $156.98 \pm 0.15$ pc \citep{radio_prior}. It has a spin frequency of 174 Hz and forms a binary system with a 0.2 \msol helium-core white dwarf \citep{Bailyn93}. The binary system has an inclination of $137.5^{\circ}$, which is favorable for measuring the Shapiro delay \citep{Shapiro64} in arriving radio pulses, which in turn provides us with an independent mass constraint. \citet{radio_prior} report a tight constraint on the mass of \jof with a value of $M = 1.418 \pm 0.044$ \msol, which is used to inform the models in this work. 

The Letter is organized as follows: In Section \ref{sec:data}, we introduce the different data sets and associated background estimates for \jof. In Section \ref{sec:modelling} we summarize the various aspects of building a pulse profile model and introduce the unique aspects required for modeling \jof, building upon the formalism described in previous NICER analyses using the pulse profile modeling code X-PSI \citep{Riley23}. Section \ref{sec:sampling} provides details of the sampling process.  We document our exploratory analyses in Section \ref{sec:exploratory analyses} where we test the effects of various modeling choices. In Section \ref{sec: Prod analyses}, we provide the results for our final model and showcase our headline results. We discuss the physical implications of the inferred model parameters by our headline model in Section \ref{sec:discussion} and conclude in Section \ref{sec:conclusions}.

\section{X-ray event data and background}\label{sec:data}

\subsection{NICER XTI} 

For this work, observations of \jof conducted by the NICER X-ray Timing Instrument (XTI) have been used to generate two different datasets. The datasets vary in the pre-processing techniques applied which are detailed in the subsections below. First we describe the common features present in both datasets.

A bright Seyfert II Active Galactic Nuclei (AGN), RX J0437.4$-$4711 \citep{HalpenMarshall96} is located within 4\arcmin.18 of our target \citep[See e.g., Figure 8 of][]{bogdanov19a}.  At this distance from the telescope aim point, $\sim 8\%$ of the AGN flux is detected, therefore providing a non-negligible contribution to the total astrophysical background.  \citet{bogdanov19a} developed an optimization technique to minimize background contamination from nearby point sources by finding the position that maximize the target's Signal-to-Noise Ratio (SNR, as defined in Equation 3 of \citealt{bogdanov19a}). For \jof, the optimal pointing is 1\arcmin.15 south-west of the pulsar, which improved the SNR by 16\% compared to directly pointing at the pulsar. With this pointing, $\sim 2.25\%$ of the AGN flux (i.e., $\sim 0.2$ counts s$^{-1}$) falls in the Field of View (FoV) of NICER and contributes to the background\footnote{We note that this percentage was estimated based on in-flight data using \texttt{nicerarf}, while the optimal pointing presented in \cite{bogdanov19a} was calculated with a pre-flight vignetting function.}. 

For our analysis, we only consider those events registered in the Pulse Invariant\footnote{Pulse here refers to the charge-pulse amplitude associated with the instrument, and not the pulsar.} (PI) channel subset [30, 300), nominally corresponding to the photon energy range 0.3--3.0 keV. The sharp low-energy threshold cutoff in the instrument calibration products coupled with the presence of increased electronic noise induced by optical light at energies below 0.3 keV can lead to biased inferences. To prevent this, we ignore the events registered below channel 30 altogether. The contribution of \jof hot regions to the counts registered above channel 300 are negligible compared to the counts generated by background processes, and are therefore also neglected. 

\begin{figure}[t]
    \centering
    \includegraphics[width=0.47\textwidth]{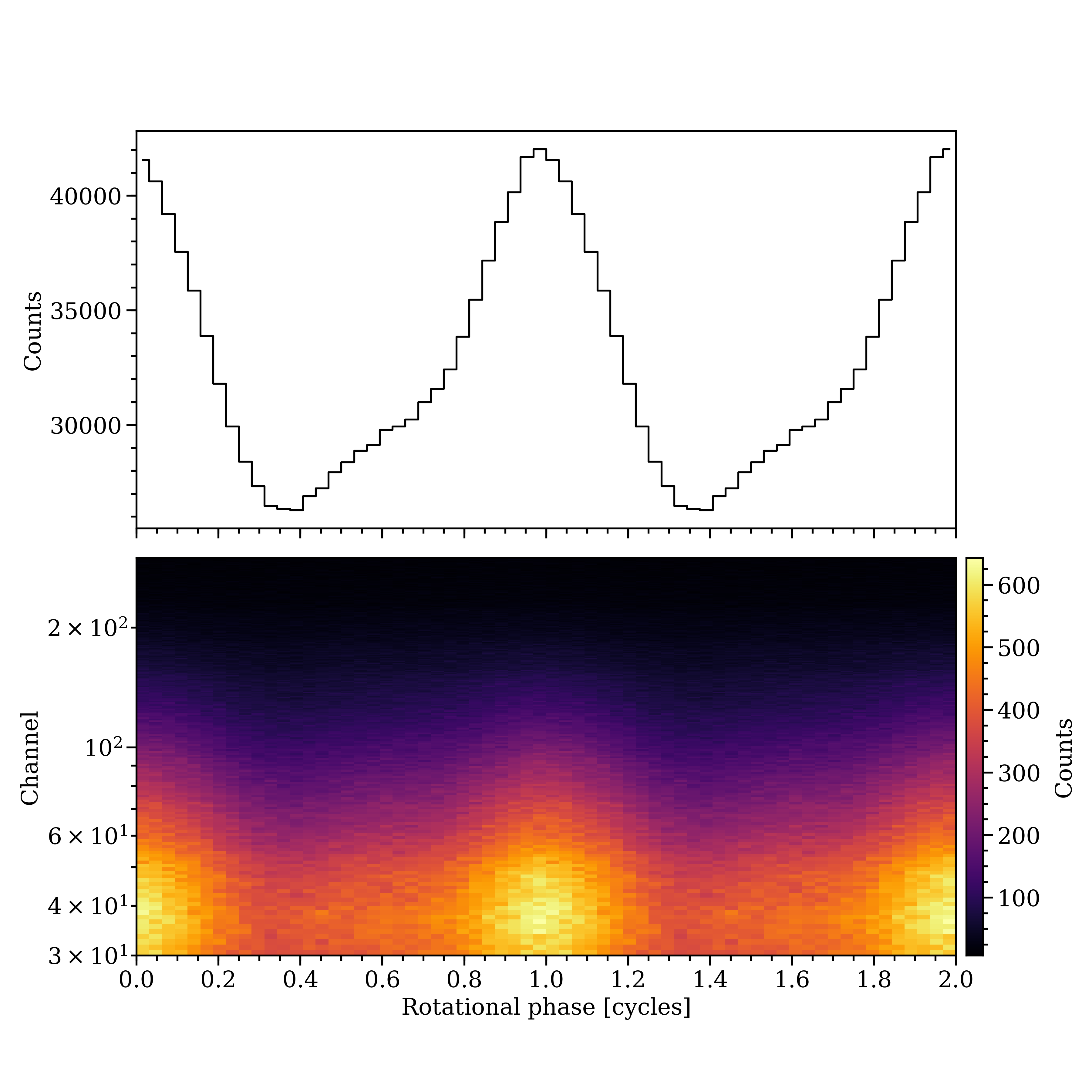}
    \caption{Phase-folded \jof (3C50) event data shown over two rotational cycles. We use 32 phase intervals (bins) per  cycle  and  the  count  numbers  in  bins  separated  by  one  cycle are identical for a given channel(s). The sum over all phase-channel pairs yields the total counts. The top panel displays the waveform obtained by summing the channels in the contiguous subset [30, 300). The bottom panel displays the phase-channel-resolved count numbers for the same channel subset, where the color bar represents counts per channel per two cycles.}
    \label{fig:J0437 count data}
\end{figure}

All the registered events accounted for are phase-folded based on the ephemeris obtained from the PPTA radio timing solutions \citep{radio_prior}, using the PINT\footnote{\url{https://github.com/nanograv/PINT}} \texttt{photonphase} tool. A corresponding pulse profile thus generated (using methods described in Section \ref{sec: 3C50 dataset})  is shown in Figure \ref{fig:J0437 count data}. Beside a prominent pulse, the profile also features a diffuse hump around phases 0.5--0.7 which \citet{Bogdanov13} suggests is a consequence of non-antipodal hot regions on the surface. This possibility is explored in our models.

\subsubsection{Delta dataset, space weather background estimate, and AGN spectrum}
\label{section: Delta dataset}

The ``Delta'' dataset (consisting of all NICER pointings in the time range 2017 July 6 to 2021 July 31). was constructed using HEASoft version 6.29c and CALDB version \texttt{xti20210707}. We first applied the standard NICER L2 processing, which applies basic filters ensuring that NICER is pointed at the target, is outside the South Atlantic Anomaly (SAA) region, and not pointed within 15$^\circ$ of the Earth limb or within 30$^\circ$ of the bright Earth, as well as standard cuts on the overshoot rate to remove time intervals with high particle background. As this analysis is sensitive to the background rate and optical loading, we applied further cuts including: (1) cutoff rigidity (\texttt{COR\_SAX}) $> 1.5$ GeV/$c$, (2) mask \texttt{DET\_ID} 34, (3) planetary K index ($K_p$) $< 5$, (4) undershoot rate $< 200$ counts s$^{-1}$. Lastly, we removed time intervals based on count rates, first by filtering out times where \texttt{DET\_ID} 14 had a count rate greater than 1.0 counts s$^{-1}$ in 8.0 s bins, then by filtering on total X-ray count rate (0.25--8.0 keV) above 6.0 counts s$^{-1}$ in 2.0 s bins. This catches times of high background that are not caught by proxies such as $K_p$ or high overshoot rate. The final good time in this dataset is 2.32 Ms.

For this dataset, we generated a background estimate using an internal version of the `Space Weather' background estimator tool\footnote{\url{https://heasarc.gsfc.nasa.gov/docs/nicer/tools/nicer_bkg_est_tools.html}}. This model uses a library of blank sky observations taken over a range of conditions of space weather ($K_p$) and the magnetic cutoff rigidity (\texttt{COR\_SAX}) which are combined to match the conditions of the observation in question. To this, a soft component representing the optical loading contamination is added based on the sun angle distribution in the observation. This background estimate represents the blank sky and does not include any contribution from contaminating sources in the field of view or non-spot emission from the pulsar.

\begin{figure}[t]
    \centering
    \includegraphics[width=0.47\textwidth]{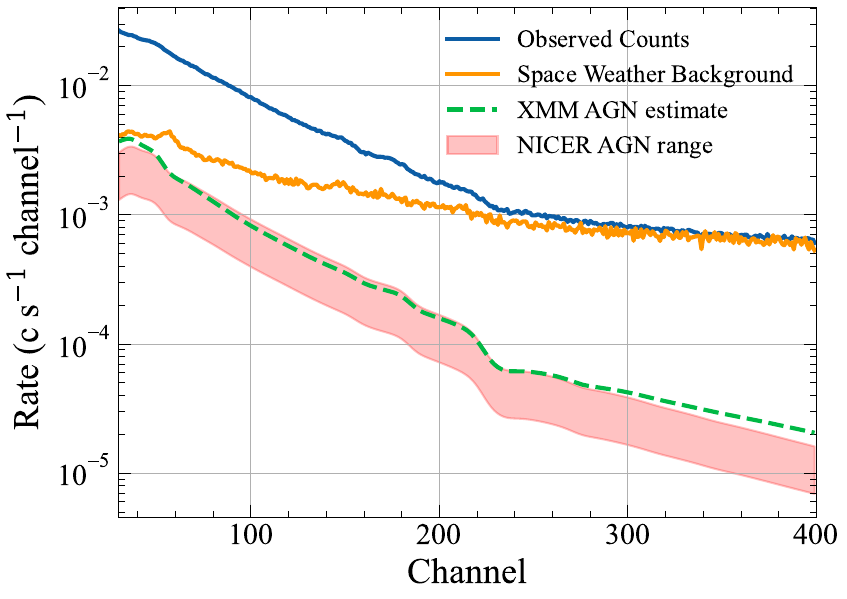}
    \caption{Count rate spectrum of the Delta dataset, along with the Space Weather background estimate and a representative range of the attenuated AGN spectra. Green is the estimated AGN spectrum for NICER based on XMM-Newton observations from 2009 and the red band shows the range of fluxes observed in NICER pointings at the AGN.}
    \label{fig:AGNspectrum}
\end{figure}

For this dataset, we also made an estimate of the contribution to the spectrum from the AGN. To do this, we created a new response matrix using the same method above but specifying the AGN position instead of the pulsar. The effective area for the AGN is a factor of 30--50 less than for the pulsar, depending on the individual detector misalignments. We constructed a model spectrum of the AGN from an absorbed double powerlaw fit to XMM-MOS1 data\footnote{We used the XMM MOS1 imaging data from ObsID 0603460101, acquired on 2009-12-15 at 19:41:41 for a total filtered exposure of 119.1~ks.} ($N_{\rm H} = 1.9(5) \times 10^{20} \;\mathrm{cm}^{-2}$\footnote{The numbers in the parentheses represent the uncertainty in the last digit.}, $\Gamma_1 = 2.9(1)$, $N_1 = 1.7(2) \times 10^{-3}$ photons/keV/cm$^2$/s, $\Gamma_2 = 1.4(1)$, $N_2 = 6.6(17) \times 10^{-4}$) photons/keV/cm$^2$/s. We also extracted spectra of the AGN from 45 NICER ObsIDs where the AGN was the target. We fit one long observation (ObsID 4060180109, 8.9 ks of good exposure) to an absorbed powerlaw ($N_{\rm H} = 1.8(4) \times 10^{20} \;\mathrm{cm}^{-2}$, $\Gamma = 2.42(5)$, $N = 1.84(6) \times 10^{-3}$ photons/keV/cm$^2$/s) and measured the flux. Using the measured XMM and NICER spectra, we constructed an estimated counts spectrum for the AGN. According to this, the AGN contributes about 0.2 counts s$^{-1}$ in the 0.3--3 keV band, which is noticeably less than the pulsar at about 1 counts s$^{-1}$. However the AGN is known to vary, so this estimate is only representative. We do not specifically know the flux state of the AGN at the time of each of the NICER observations of the pulsar. Figure \ref{fig:AGNspectrum} shows the observed count rate spectrum from the Delta dataset, along with the Space Weather background estimate and and the attenuated AGN spectra. The NICER spectrum is shown as a range covering the lowest to highest fluxes observed in the individual NICER ObsIDs.

\subsubsection{3C50 dataset, background model, and AGN spectrum}
\label{sec: 3C50 dataset}

The Delta dataset is processed to minimize background contamination and provide a space weather estimate. However, this estimate does not have any quantifiable statistical and systematic uncertainty associated with it. For our analysis using the space weather estimate, we therefore only use it to place a conservative lower limit on the background (see Section \ref{sec:sampling}).

A substantial portion of the non-source background in a typical NICER observation originates from the local environment of the telescope. The NICER team has developed several approaches for modeling this time-dependent background emission. One is the 3C50 model \citep[based on a library of blank fields,][]{Remillard22}, which makes use of the NICER event data for the source of interest to produce an estimate of the background level as a function of energy.

The 3C50 dataset is based on a particular list of Good Time Intervals (GTIs) that pass filtering tests designed to diagnose when the 3C50 background model is performing optimally. The 3C50 model is defined in \cite{Remillard22}, and the analysis and filtering procedures are described in detail in \cite{Salmi22}. The filtering steps reduce the usable exposure time, but this method provides access to quantified uncertainties in the net background spectrum, which is dominated by systematic effects that are otherwise difficult to determine.  The investigation of seven rotation-powered X-ray pulsars which have NICER archival exposures of 1--3 Ms revealed that the root-mean-square (rms) deviations in the net source intensity (0.3--2.0 keV), when the data is binned at 200--400 ks, is very close to 0.018 counts s$^{-1}$ ($1 \sigma$) for each pulsar, regardless of the average pulsar intensity \citep{Salmi22}.  Given the defensible assumption that these pulsars are stable in X-ray intensity, the common rms values are interpreted as a measure of the systematic uncertainty for the 3C50 background model, for filtered exposure times of 200 ks or longer. Furthermore, it was found that the rms deviations in net intensity, in different energy bands, are proportional to the average background rate in the same bands. This implies that the deviations are tied to the precise level of the average background spectrum, and it enables estimation of the background uncertainty in any energy band.  The 3C50 filtering effort therefore constrains pulse fractions, to some degree, in the effort to physically model the pulsar emission. One caveat, here, is the possible presence of additional ``background'' components, in the form of extra X-ray sources in the NICER FoV.  Such considerations have to be made for each target, and they are particularly important in the case of NICER investigations of \jof.

The procedures to define GTIs and to filter results for the 3C50 dataset \citep{Salmi22} are briefly summarized, as follows.  Exposure intervals for a given target are first determined with the HEASoft tool \texttt{nimaketime}, and the times are masked by $\pm 30$ s at any sunshine on/off boundary.  The remaining continuous intervals, each of duration $T_i $, are selected if $T_i \ge 200 $ s, and they are divided into $N$ equal pieces if $T_i > 450 $ s, where $N = int(T_i / 300 + 0.5)$. These steps provide a list of GTIs with an average exposure $\sim$ 300 s and exposure range 200--450 s.

For each GTI, the raw NICER spectrum is extracted, a background prediction is obtained with the 3C50 model, and then six filtering tests are performed on the results. Two filters apply to parameters needed for the background model: $nz < 220$ counts s$^{-1}$, where $nz$ is the noise rate normalized to 50 Focal Plane Modules (FPMs) at 0.0--0.25 keV, and $ibg < 0.2$ counts s$^{-1}$, where $ibg$ is the normalized rate of good events at 15--18 keV (beyond the effective area limits of the optics). The next two filters \citep[as described in ][]{Remillard22} operate on the net background-subtracted spectrum, $S0_{\rm net} < 0.15$ counts s$^{-1}$ and $hbg_{\rm net} < 0.05$ counts s$^{-1}$, where the corresponding energy bands are 0.2 to 0.3 keV for $S0$ and 13--15 keV for $hbg$. Finally, when the net pulsar spectrum shows extremely faint or undetectable intensity above 2.0 keV, we add filters in the C band (2-4 keV) and D band (4-12 keV): $C_{\rm net} < 0.1$ counts s$^{-1}$ and $D_{\rm net} < 0.3$ counts s$^{-1}$.

The dataset for \jof considered in this letter consists of all NICER pointings in the time range 2017 July 6 to 2021 October 11.  The raw exposure (\texttt{nimaketime}, with geometric filters, but no rate filters) is 2.736 Ms, and the yield is reduced to 1.328 Ms after all of these filtering steps are applied (therefore, lower than the final good time of the Delta dataset). Therefore the total number of counts in the final 3C50 dataset (2095698 counts) is lower than that of the Delta dataset (3863743 counts).  In addition, 1.310 Ms of these GTIs uniformly utilize 50 FPMs (excluding only 14 and 34), and we sidestep the need to normalize the input data and vary the response files, for each GTI, by adopting the 50-FPM subset (4442 GTIs; 1.310 Ms) as the 3C50 dataset (see also Section \ref{sec: instrument model}).  In the 3C50 dataset, the background-subtracted average intensity at 0.3--2.0 keV is 1.079 counts s$^{-1}$.  Pertinent to filtering in the C and D bands, the net intensities there are 0.020 and -0.004 counts s$^{-1}$, respectively.

As noted previously, the NICER observations of PSR J0437--4715 contain a known AGN, RX J0437.4$-$4711, near the edge of the FoV. Contributions from this source can be regarded as an additional background component, with its own uncertainty. NICER conducted separate, on-axis observations of RX J0437.4$-$4711 during 2017 July 10-20, 2017 October 8-10, 2021 December 10-26, and 2022 February 28 - March 19.  The mean spectrum for the AGN was determined for each epoch, using the 3C50 background model and level 2 filtering applied to each GTI \citep{Remillard22}. The average intensity (0.3--3.0 keV) is 6.26 counts s$^{-1}$, and the rms deviation between the four epochs is 25\%. On the other hand the response calculator for NICER (\texttt{nicerarf}) determines that the AGN offset position in the primary observations of pulsar \jof, integrated over the GTIs in the \jof dataset, has an effective area reduced by a factor of 42. The expected contribution from the AGN in the pulsar dataset is then represented by the mean AGN spectrum, rescaled by a factor of 42. The uncertainty range for this component is conservatively estimated as -50\% to +100\% of the values in this spectrum, based on NICER and XMM-Newton observations.

\subsection{XMM-Newton EPIC}
\label{sec: XMM}

The XMM-Newton European Photon Imaging Camera (EPIC) event dataset used in this analysis was extracted from two archival observations of \jof acquired on 2002 October 09 (ObsID 0112320201) and 2009 December 15 (ObsID 0603460101), with total unfiltered on-source exposures of 69.4 and 129.4 ks, respectively.  In this work, we only make use of the `Full Frame' imaging mode data from the EPIC MOS1 and MOS2 instruments \citep{Turner2001}. Although lacking high time resolution, they provide reliable low background phase-averaged source spectra desirable for the analyses described below.  While the EPIC pn data acquired in `Timing' mode allows the study of the pulsed emission, it is not used here due to the considerably larger uncertainties in the calibration of the instrument in this observing mode compared to MOS1 and MOS2\footnote{For details, see the XMM-Newton calibration documentation at \url{https://www.cosmos.esa.int/web/xmm-newton/calibration}}. 

The source event list extraction of the EPIC MOS data were conducted with the Science Analysis Software (SAS\footnote{The \xmm SAS is developed and maintained by the Science Operations Centre at the European Space Astronomy Centre and the Survey Science Centre at the University of Leicester.}) version \texttt{xmmsas\_20211130\_0941}. Following the same filtering procedure as in \citet{Bogdanov13}, the event data were screened for instances of high particle background count rates and the recommended PATTERN ($\le$12 for MOS1/2) and FLAG ($0$) filters were applied. This resulted in 182.5 and 182.8 ks of clean effective exposure for the MOS1 and MOS2, respectively. The X-ray count rates of the pulsar in the cleaned event data from the two epochs are consistent within uncertainties, as expected from the steady thermal emission from rotation-powered MSPs.  We extracted source and background counts from circular regions of radius 32$''$ and 105$''$, respectively, to produce spectra in the form of counts per instrument channel. The SAS command \texttt{backscale} was then used to calculate the area of source and background regions used to extract the spectral files\footnote{See \url{https://www.cosmos.esa.int/web/xmm-newton/sas-threads}}. The value of this parameter is written into the header of the SPECTRUM table of the file as keyword BACKSCAL.

\section{Pulse Profile Modeling Using X-PSI}\label{sec:modelling}

We make use of the \ac{X-PSI} software package \citep{Riley23}\footnote{\url{https://github.com/xpsi-group/xpsi}} to develop models and perform Bayesian parameter estimation by analyzing the X-ray emission observed from \jof. The core modeling procedure is extensively described in \citet{Riley19, Riley21, Bogdanov19b, Bogdanov21}. We summarize the process in this section and detail the novel components required in this work. Our analyses span \ac{X-PSI} versions \texttt{v0.7.0} and \texttt{v2.2.2}, and the exact versions used in each of the inference runs can be found in the output files in the Zenodo repository\footnote{\url{https://doi.org/10.5281/zenodo.10886504}}. The analysis files can be found on the Zenodo repository including the data products, the numeric model files, scripts, and Jupyter notebooks in the Python language to produce the results and figures of this article using the X-PSI framework, and posterior sample files.

\subsection{Radiation from source to telescope}
\label{subsec: radio priors}

\ac{X-PSI} performs relativistic ray-tracing of the neutron star surface emission using the Oblate Schwarzschild plus Doppler approximation \citep{Poutanen03,Poutanen06,Cadeau07,Morsink07,AlGendy14,Bogdanov19b}. We define a flat prior density distribution on the equatorial radius ($R_{\rm eq}$), which is then shaped by limits imposed on the compactness and surface effective gravity as described in section 2.2.1 of \citet{Riley21}.  

Radio-frequency timing of \jof provides precise measurements for some of our model parameters. The pulsar mass ($M$), distance ($D$), and orbital inclination angle ($i$) can be measured with much higher precision with this technique than with the X-ray observations. We therefore inform our models using the latest pulsar timing analyses from the PPTA, described in \citet{radio_prior}. Specifically, we adopt the PPTA measurements of $M$, $D$, and $i$ as Gaussian-distributed prior Probability Density Functions (PDFs) for our Bayesian inference.

The pulsar distance is measured through the Shklovskii effect \citep{Shklovskii70} on the orbital period ($P_b$) of the pulsar, which induces an apparent time derivative, $\dot{P}_b$. The measured $\dot{P}_b$ is corrected for contributions from differential acceleration in the Galaxy (circular and Galactic potential) and for the small contribution due to gravitational wave emission from the system. The resulting distance constraint is $D = 156.98\pm0.15$\,pc \citep{radio_prior}, which is consistent with independent (but more uncertain) distances derived from other methods.

The orbital inclination is derived from a measurement of annual-orbital parallax \citep{Kopeikin95}. This effect is now particularly useful because precise independent measurement of $i$ breaks the strong covariance between the range (i.e., companion mass) and shape ($\sin i$) parameters that describe the Shapiro delay. Therefore, the measured Shapiro delay primarily informs the companion mass ($M_c$), and a precise constraint for $M$ then follows from the binary mass function. From these effects, \citet{radio_prior} derive $i = 137.506 \pm 0.016$\,degrees and $M_p = 1.418 \pm 0.044$\,M$_\odot$ (where the uncertainties represent the standard 68\% confidence interval). The parameters can be considered uncorrelated.

For our final fiducial results, we use priors corresponding to the values quoted above, which were derived from the recent PPTA third data release (PPTA-DR3) and its noise analysis \citep{ppta-dr3, ppta-dr3-noise}. However, model selection for the hot spot geometry (Section \ref{sec:exploratory analyses}) was conducted using a slightly different prior for these parameters, which was derived from an earlier generation of PPTA data, described as PPTA-DR2.5 in \citet{radio_prior}. The values used for these earlier analyses were $D = 156.69\pm0.30$\,pc, $i =137.485 \pm 0.038 $\,degrees, and $M_p = 1.411 \pm 0.056$\,M$_\odot$, which are highly consistent with the updated PPTA-DR3 measurements. 

The interstellar medium along the line of sight of the pulsar can get ionized by photoabsorption of the emergent soft X-rays thereby attenuating the signal generated by the source. The attenuation factor at any given energy is dependent on the neutral  hydrogen column density $N_{\rm H}$, which forms a parameter in our models. The relative abundances from \citet[][updated in 2016]{Wilms2000} are adopted to compute tables, using the \texttt{tbnew}\footnote{\url{https://pulsar.sternwarte.uni-erlangen.de/wilms/research/tbabs/}} model, relating the attenuation factor to energy for a given $N_{\rm H}$. For further details, we refer the reader to section 2.4 of \citet{Bogdanov21}.

Various estimates are available for the $N_{\rm H}$ along the line of sight of \jof. Employing the relation between Dispersion Measure (DM) and $N_{\rm H}$ from \citet{He13}, along with the DM measurement reported by \citet{radio_prior}, yields $N_{\rm H} \approx 0.8\times 10^{20} \rm cm^{-2}$. The HEASARC mapping tool\footnote{\url{https://heasarc.gsfc.nasa.gov/cgi-bin/Tools/w3nh/w3nh.pl}} also arrives at the same value. Estimates based on 3D \textit{E (B -- V)} extinction maps (E (B -- V)$\sim 0.002 \pm 0.014$ using Stilism\footnote{\url{https://stilism.obspm.fr}}) \citep{Lallement14, Capitanio17}, coupled with the relation between extinction and $N_{\rm H}$ provided by \citet{Foight16}, yield $N_{\rm H} \approx 0.2\times 10^{20} \rm cm^{-2}$. Finally, spectral fits of UV and X-ray data to \jof by \citet{Gonzalez19} yield a range of $0.7 - 2.4 \times 10^{20} \rm cm^{-2}$ for the  $N_{\rm H}$. 

Based on these results, we initially tested different models with a wide and uniform prior $\mathcal{U}(0.01, 5.0)\times 10^{20} \rm cm^{-2}$, and always inferred rather low values. Therefore, for our later runs, we opted for a narrower prior of $\mathcal{U}(0.004, 2.0)\times 10^{20} \rm cm^{-2}$ to better sample the lower tail.

\subsection{Surface emission}

Since it is essentially impossible to spatially resolve a neutron star, we have to rely on phenomenological models of the surface radiation field, described in this section, to reproduce the observed X-ray event data. These models include sufficient complexity such that they replicate our basic understanding of possible magnetic field configurations, wherein the magnetospheric return currents deposit energy on the surface and heat it up \citep[see e.g.][]{Harding01}. Note that internal heating by magnetic field decay is not significant for old and low field pulsars such as \jof and other MSPs \citep{Gonzalez10}.

\subsubsection{Hot spot configurations}
\label{subsec: HR config}

Following the prescription laid down in \citet{Riley19}, we model the hot regions using (potentially compound) spherical caps on the neutron star surface. We start with simplistic models and increase complexity based on model performance, also checking for complexities to which the likelihood function is insensitive and is hence unhelpful. 

All our models involve two hot regions that are barred from overlapping with each other and are referred to as {\it primary} and {\it secondary} spots. Schematics of our various models can be found in  figure 1 of \citet{Vinciguerra23a} . We report our analysis for the following increasingly complex models \footnote{In our preliminary analyses on an initial dataset not reported in this letter, we also tested \texttt{ST-S}, \texttt{CST-U}, \texttt{ST+PST} (also tested on the Delta dataset), and \texttt{PST-U} configurations not included in this list. \texttt{ST-S} was utterly incapable of explaining the data, resulting in large residual structures. \texttt{CST-U} and \texttt{ST+PST} only led to very nominal model performance improvements over \texttt{ST-U}. \texttt{PST-U} was plagued by multiple degenerate modes and the sampler could not converge in a reasonable timescale.}\textsuperscript{,}\footnote{We refer the reader to section 2.5 of \citet{Riley19} and to sections 2.2 and 2.3.4 of \citet{Vinciguerra23a} for the full prior definitions of these models.}:

\begin{itemize}
    \item {\bf  ST-U}: The Single Temperature -- Unshared (\texttt{ST-U}) model consists of two disjoint simply-connected spherical caps of uniform effective temperature for the atmosphere. In this particular model, the spot with the lower colatitude is designated as the primary. This condition aside, {\it unshared} here implies that the parameters of one spot are independent of the other.
    \item {\bf ST+PDT}: Alongside the primary \texttt{ST} spot, we have a secondary Protruding Dual Temperature (\texttt{PDT}) spot consisting of two spherical overlapping emitting components. In the overlapping portion, emission of only one of the components is accounted for, and that is referred to as being the {\it superseding} component.
    \item {\bf CST+PDT}: In this model, we have a Concentric Single Temperature (\texttt{CST}) spot as our primary, consisting of two spherical components, one emitting and one \textit{masking}, thus forming a ring. The \textit{masking} component (referred to as the \textit{omitting} component in \citet{Vinciguerra23a, Vinciguerra23b}) is ensured to be concentric within the emitting component. The secondary spot here is a \texttt{PDT}.
\end{itemize} 

Although further complexities can be introduced, we end our model sequence with \texttt{CST+PDT} given its ability to sufficiently explain the data (as explained in Section \ref{sec: Prod analyses}) and the computational expense that would be incurred upon thoroughly exploring more complex models.

\subsubsection{Atmosphere}

All emitting hot region components incorporate a geometrically thin, fully-ionized hydrogen atmosphere model to describe the specific intensity of the local comoving effective temperature field. The intensity is interpolated over lookup tables of precomputed values, generated using \texttt{NSX} \citep{HoandLai01}, for a given effective temperature, surface gravity, photon energy and cosine of the emission angle (see Section 2.4.1. of \citet{Riley19} for more details). Compared to \citet{Riley19}, this work utilizes an extended version of the lookup tables, as was done in all the later publications that used \ac{X-PSI}. The limitations and uncertainties that this choice of atmosphere model implies are discussed in detail in \citet{Bogdanov21, Salmi23}.

\subsection{Instrument response models}
\label{sec: instrument model}

For developing the NICER instrument response models, we utilize tailored Ancillary Response Files (ARFs) and Redistribution Matrix Files (RMFs) to generate an off-axis response matrix corresponding to a particular dataset. The off-axis pointing induces projection effects on the instrument's effective area, which is nearly energy independent, reducing it by an average factor of $\sim 0.92$ over all channels.

For the Delta dataset, we created custom response matrices using the \texttt{nicerarf} and \texttt{nicerrmf} tools, specifying the pulsar position rather than the XTI pointing direction. This produces an ARF that is corrected for the vignetting and time-variable number of active detectors. The product of the ARF and RMF is generated as a single response matrix (RSP) file.
The 3C50 dataset uniformly selects the same 50 FPMs for each GTI (Section \ref{sec: 3C50 dataset}). A single pair of response files, computed with \texttt{nicerarf} and \texttt{nicerrmf}, are used for this dataset, combining 50 of the 52 FPMs, while excluding FPMs 14 and 34.

The XMM-Newton EPIC MOS1 and MOS2 RMF and ARF products were generated using the \texttt{rmfgen} and \texttt{arfgen} tools in SAS. These resulting response files, along with the source and background spectral files, for the two archival observations were merged into a single set using a weighted average based on the relative exposures of the two observations.

To capture the $\sim \pm 10$ \% calibration uncertainty associated with the effective area (and assuming that this is energy-independent), we define an absolute energy-independent instrument-scaling factor $\alpha_{\rm XTI}$ that operates on the response matrix. In some of the earlier Delta runs, we combined this with the pulsar distance $D$ [kpc] into a single flux-scaling variable $\beta\,[\mathrm{kpc^{-2}}] = \alpha_{\mathrm{XTI}} D^{-2}$ that we are sensitive to. This choice reduces the number of free parameters to be sampled and helps save on computational resources.  This parameterization is further detailed in \citet{Vinciguerra23a}. For the later runs, we fix the distance altogether considering its highly informative prior and only sample $\alpha_{\rm XTI}$.

When we jointly fit for NICER and XMM-Newton data, we introduce another energy-independent effective area term $\alpha_{\rm MOS}$ that is applied to the responses of the XMM-Newton cameras, MOS1 and MOS2. $\alpha_{\rm XTI}$ and $\alpha_{\rm MOS}$ share a correlated prior PDF as explained in sections 2.4 and 4.2 of \citet{Riley21}.  We apply a shared scaling uncertainty factor of 10\% and telescope-specific uncertainty factors of 3\% leading to a global 10.4\% uncertainty in the overall scaling factors.

\subsection{X-PSI Settings}
\label{sec: X-PSI settings}

Modeling in \ac{X-PSI} requires specification of certain resolution settings related to the hot region cell mesh discretisation, ray emanation, energy and temporal hyperslice discretisation (for stellar rotation), and multiple imaging. We specify the default values of the \ac{X-PSI} settings\footnote{\label{footnote: xpsi settings}{For further details on the settings specified in Table \ref{tab:x-psi settings}, please visit: \url{https://xpsi-group.github.io/xpsi/hotregion.html}}} that we use for all runs in Table \ref{tab:x-psi settings}, unless explicitly stated otherwise\footnote{For certain runs, we experiment with the different resolution settings as a crude check on inference sensitivities.}. 

\begin{table}
    \centering
    \begin{tabular}{c|c|c} \hline 
         \textbf{X-PSI setting}& \textbf{Default value} &\textbf{Headline value}\\ \hline 
         \hline
         \texttt{sqrt\_num\_cells}& 32 &18\\ \hline 
         \texttt{min\_sqrt\_num\_cells}& 10 &10\\ \hline 
         \texttt{max\_sqrt\_num\_cells}& 80 &32\\ \hline 
         \texttt{num\_leaves}& 64 &32\\ \hline 
         \texttt{num\_energies}& 128 &64\\ \hline 
         \texttt{num\_rays}& 512 &512\\ \hline 
         \texttt{image\_order\_limit}& 3 &3\\ \hline
    \end{tabular}
    \caption{Default and headline-result X-PSI settings\footref{footnote: xpsi settings} used in this work. The default settings are used unless explicitly stated otherwise. Some of the headline values are smaller than the default values to compensate for computational expense incurred by improved sampling of the most complex model in our analysis.}
    \label{tab:x-psi settings}
\end{table}

\section{Sampling process}\label{sec:sampling}

\subsection{Likelihood function and background}
\label{subsec: likelihood}

In this section, we briefly describe the likelihood function formulation for the different background constraints as applied to various models in this letter. The origin and consequences of various assumptions and choices briefly mentioned here are elaborated in \citet{Riley19}, \citet{Riley21} and \citet{Salmi22}.

Our framework involves a background model aimed at capturing any background emission external to the neutron star. These are expected to be registered as homogeneous Poissonian events, which we attempt to quantify by defining a set of statistically independent background count-rate (nuisance) variables $\{\mathbb E[b]\}$, operating one per channel. This entails the notion that background emission processes, unlike the neutron star signal, will not coherently sum up with the rotation period of the pulsar upon phase-folding and will hence be subsumed by the phase-invariant background model. 

Our likelihood function is therefore conditional on our regular model parameters and the background count-rate parameters. To save on computational expense, we numerically marginalize this likelihood function over these nuisance parameters \citep[see equations 4 and 9 in][]{Riley19}. We specify channel-separable flat prior PDFs between a given range, i.e., $\mathbb E[b] \sim U(\mathcal{L},\; \mathcal{U})$, where $\mathcal{L}$ and $\mathcal{U}$ are lower and upper bounds for the prior support at a given channel. When our analyses involve multiple instruments, we have separate background models (and associated prior supports). Here, we denote the NICER background model as $\mathbb E[b_{\rm N}] \sim U(\mathcal{L_{\rm N}},\; \mathcal{U_{\rm N}})$ and the XMM-Newton background model as $\mathbb E[b_{\rm X}] \sim U(\mathcal{L_{\rm X}},\; \mathcal{U_{\rm X}})$. 

In this work, for the NICER-only analyses, when we do not have any specific information on the upper background limit, the upper bound of the prior support is left unspecified \citep[see Appendix B.2.3.2 of][for more details]{Riley_thesis}. When we use the Delta dataset and the space weather estimate, we specifically define only a lower limit as $\{\mathcal{L_{\rm N}}\} := 0.9 \times \{\mathcal{B_{\rm {N,\; SW}}}\}$, where $\{\mathcal{B_{\rm {N,\; SW}}}\}$ is the set of NICER background count rates from the space weather estimate.  

When using the 3C50 dataset and background estimate in combination with either of the AGN spectra (Delta or 3C50), we apply them in three different ways. We can choose to specify only a lower instrument background limit\footnote{By instrument background, we refer to all processes that register events not originating from the \jof FoV but from within the Solar System instead.} as $\{\mathcal{L_{\rm N}}\} := {\rm max}(0,\; \{\mathcal{B_{\rm {N,\; 3C50}}}\} - n_{l}\{\sigma_{\mathcal{B_{\rm {N,\; 3C50}}}}\})$, where $\{\mathcal{B_{\rm {N,\; 3C50}}}\}$ is the set of NICER background count rates from the 3C50 instrument background estimate, $\{\sigma_{\mathcal{B_{\rm {N,\; 3C50}}}}\}$ is the total standard deviation\footnote{The total standard deviation here is square root of the quadratic sum of the systematic and statistical errors \citep[See Section 3.4 and Appendix A.2 of ][for more details.]{Salmi22}}, and $n_{l}$ is the degree of conservatism acting on the lower standard deviation. We can also specify only a lower limit consisting of the instrument background estimate and an AGN spectrum as $\{\mathcal{L_{\rm N}}\} := {\rm max}(0,\; \{\mathcal{B_{\rm {N,\; 3C50}}}\} - n_{l}\{\sigma_{\mathcal{B_{\rm {N,\; 3C50}}}}\} + 0.25\times\{\mathcal{B_{\rm {N,\; AGN}}}\})$, where $\{\mathcal{B_{\rm {N,\; AGN}}}\}$ is the set of AGN count-rate observed by NICER. We sometimes specify an upper limit on the background consisting of the instrument background estimate and an AGN spectrum as  $\{\mathcal{U_{\rm N}}\} := \{\mathcal{B_{\rm {N,\; 3C50}}}\} + n_{u}\{\sigma_{\mathcal{B_{\rm {N,\; 3C50}}}}\} + 2\times\{\mathcal{B_{\rm {N,\; AGN}}}\}$, where $n_{u}$ is the degree of conservatism acting on the upper standard deviation. Considering the variability of the AGN, the above range of 0.25--2 $\times$ the AGN estimate is chosen to encapsulate the observed historic maximum and minimum of the AGN X-ray flux (see Figure \ref{fig:AGNspectrum} and the discussion in Section \ref{sec: 3C50 dataset}). 

When involving XMM, the corresponding lower and upper instrument background limits are set as  $\{\mathcal{L_{\rm X}}\} := {\rm max}(0,\; \{\mathcal{B_{\rm {X}}}\} - n\sqrt{\{\mathcal{B_{\rm {X}}}\}})$ and $\{\mathcal{U_{\rm X}}\} := \{\mathcal{B_{\rm {X}}}\} + n\sqrt{\{\mathcal{B_{\rm {X}}}\}}$, where  $\{\mathcal{B_{\rm X}}\}$ is the set of XMM-Newton background count numbers based on blank-sky observation, and $n=4$ is the chosen degree of conservatism (as was done in \citet{Riley21, Salmi22}).

The exact applications of the different lower and upper limits in the different models are detailed in Sections \ref{sec:exploratory analyses} and \ref{sec: Prod analyses}.

\subsection{Posterior computation}

\begin{table}
    \centering
    \caption{Default and headline-result \Multinest settings used in this work. The default settings are used unless explicitly stated otherwise.}
    \label{tab: MultiNestParams}
    \begin{tabular}{c|c|c} \hline 
         \textbf{\Multinest setting}& \Centerstack{\textbf{Default}\\ \textbf{value}} &\Centerstack{\textbf{Headline}\\ \textbf{value}}\\ \hline 
         \hline
         Sampling Efficiency (SE)& 0.3 & 0.3\\ \hline 
         Evidence Tolerance (ET)& 0.1 & 0.1\\ \hline 
         Live Points (LP)& 4,000 & 20,000\\ \hline 
         Multi-mode (MM)& off & off\\ \hline
    \end{tabular}
    \parbox{\columnwidth}{\justifying \tablecomments{Reducing the values of sampling efficiency and evidence tolerance, and increasing the number of live points leads to enhanced accuracy in the calculation, albeit at an increased computational expense.
    The multi-mode variant locks in a select number of live points onto identified modes, the number of which depends on its prior mass, in order to better explore them. This, however, slightly compromises the precision of the final results, assuming the same settings otherwise.}}
\end{table}

Once we construct the likelihood function and define our prior PDFs using X-PSI, we employ \Multinest \citep{Feroz09}, interfaced via \Pymultinest \citep{Buchner14}. \Multinest uses a Bayesian inference technique known as nested sampling \citep{Skilling04} that targets estimation of the evidence. This requires exploration of the parameter space, which in turn yields posterior distributions as a byproduct. Table \ref{tab: MultiNestParams} summarizes the most relevant settings\footnote{For further details on the settings specified in Table \ref{tab: MultiNestParams}, we refer the reader to section 2.4 of \citet{Vinciguerra23a}.} pertaining to our analyses and their default values that we use for most of our exploratory runs, unless explicitly stated otherwise. Our two \texttt{CST+PDT} headline runs use a much higher resolution of 20,000 live points for thorough exploration of the parameter space. 

\section{Exploratory analyses}
\label{sec:exploratory analyses}

In our analyses, we performed a number of exploratory runs using the \texttt{ST-U} and \texttt{ST+PDT} models and the PPTA-DR2.5 radio priors described in Section \ref{subsec: radio priors}. We tested how the different dataset choices, spot patterns, background models, and runtime settings affect our inferences. In this section, we document the results of these exploratory runs. Thus, having identified the important factors, and via model comparison and quality checks, we arrive upon \texttt{CST+PDT} as our preferred model for \jof, the results of which are described in Section \ref{sec: Prod analyses}. We performed production quality runs for the \texttt{CST+PDT} model which constitute our headline result. We summarize all the runs reported in this letter in Table \ref{tab: run summary}.

\begin{table*}
    \centering
    \begin{tabular}{c|c|c|c} \hline 
         \textbf{Hot region model}&  \textbf{Dataset}&  \textbf{Lower background limit}& \textbf{Upper background limit}\\ \hline 
         \hline
         \texttt{ST-U + elsewhere}& Delta&  Space weather& -\\ 
         \hline
         \multirow{12}{*}{\texttt{ST-U}}
         &  Delta&  -& -\\ \cline{2-4} 
         &  Delta + XMM& from XMM&from XMM\\ \cline{2-4} 
         &  3C50&  -& -\\ \cline{2-4} 
         &  3C50& 3C50 BKG: $n_l = 1$& -\\ \cline{2-4} 
         &  3C50& 3C50 BKG: $n_l = 2$& -\\ \cline{2-4} 
         &  3C50& 3C50 BKG: $n_l = 3$& -\\ \cline{2-4} 
         &  3C50& 3C50 BKG: $n_l = 4$& -\\ \cline{2-4} 
         &  3C50& 3C50 BKG: $n_l = 3$ (smoothed)& -\\ \cline{2-4} 
         &  3C50&  \Centerstack{3C50 BKG: $n_l = 3$ (smoothed) \\ + $0.5 \times $Delta AGN estimate \\ (scaled to 3C50 exposure)}& -\\ \cline{2-4} 
         & 3C50& \Centerstack{3C50 BKG: $n_l = 3$ (smoothed) \\ + $0.5 \times $Delta AGN estimate\\ (scaled to 3C50 exposure)}&\Centerstack{ 3C50 BKG: $n_u = 3$ (smoothed) \\ + $2\times $Delta AGN estimate \\ (scaled to 3C50 exposure)}\\ \cline{2-4} 
         \hline
         \multirow{5}{*}{\texttt{ST+PDT}}
         & Delta&  Space weather& -\\ \cline{2-4} 
         & 3C50& -& -\\ \cline{2-4}
         & 3C50& 3C50 BKG: $n_l = 3$ (smoothed)& -\\ \cline{2-4} 
         & 3C50& \Centerstack{3C50 BKG: $n_l = 3$ (smoothed) \\ + $0.5 \times $3C50 AGN estimate}&\Centerstack{3C50 BKG: $n_l = 3$ (smoothed) \\ + $2 \times $3C50 AGN estimate}\\ \cline{2-4} 
         \hline
         \multirow{6}{*}{\texttt{CST+PDT}}
         & 3C50& -& -\\ \cline{2-4}
         & 3C50& 3C50 BKG: $n_l = 3$ (smoothed)& -\\ \cline{2-4} 
         & 3C50& \Centerstack{3C50 BKG: $n_l = 3$ (smoothed) \\ + $0.5 \times $3C50 AGN estimate}&\Centerstack{3C50 BKG: $n_l = 3$ (smoothed) \\ + $2 \times $3C50 AGN estimate}\\ \cline{2-4} 
         & 3C50 + XMM& \Centerstack{3C50 BKG: $n_l = 3$ (smoothed) \\ + $0.5 \times $3C50 AGN estimate, \\ and from XMM}&\Centerstack{3C50 BKG: $n_l = 3$ (smoothed) \\ + $2 \times $3C50 AGN estimate, \\ and from XMM}\\ 
         \hline
         \multirow{3}{*}{\Centerstack{\texttt{CST+PDT} \\ (High-resolution)}}
         & 3C50& -& -\\ \cline{2-4}
         & 3C50& \Centerstack{3C50 BKG: $n_l = 3$ (smoothed) \\ + $0.5 \times $3C50 AGN estimate}&\Centerstack{3C50 BKG: $n_l = 3$ (smoothed) \\ + $2 \times $3C50 AGN estimate}\\ \cline{2-4}
         \hline
         \Centerstack{\texttt{CST+PDT} \\ (No radio priors)}& 3C50& -& -\\
         \hline
    \end{tabular}
    \caption{Summary of datasets used and background implementations for all the runs listed in Sections \ref{sec:exploratory analyses} and \ref{sec: Prod analyses}.}
    \label{tab: run summary}
\end{table*}

\subsection{ST-U}
\label{sec: ST-U results}

\subsubsection{Effect of datasets}
\label{sec: ST-U datasets}

The \texttt{ST-U} model was initially run on the Delta dataset. Given that this is the simplest and most computationally inexpensive model in our list, we also tested the model on the 3C50 dataset, once that was made available. Besides bench-marking the effects of switching to a new dataset, we also use this model to test the possible implementations of the associated background constraints. 

To assess model consistency between the two datasets, we opt to compare the runs involving the fewest differences between them.
The inference runs on the Delta dataset were performed both with and without the space weather estimate. However in the first case, our model also included the X-PSI \texttt{elsewhere} component, which allowed for the rest of the star (outside of the hot regions) to emit blackbody radiation with a singular uniform effective temperature. 
The runs on the 3C50 dataset were performed for a variety of cases, namely: without any background constraints; with only lower instrument-background constraints; with only lower background constraints involving both instrument and AGN estimates; and finally with both lower and upper background constraints. The details of these limits are specified in Section \ref{subsec: likelihood}. 
The closest comparable runs between the Delta and 3C50 datasets are therefore the ones involving no background limits whatsoever.

Before comparing the two runs, we note the following differences between them. The Delta run uses all of our default \Multinest settings while all the 3C50 runs for this model use lower resolution settings consisting of half the number of live points, i.e., LP 2000 and a SE of 0.8, to save on computational resources for testing purposes. During some early tests on an initial dataset\footnote{Not included in this letter.} , we found that the inferred results and model performance were virtually identical upon increasing the live points from 2000 to 4000. 
As for the \ac{X-PSI} resolution settings, the Delta run uses a lower number for \texttt{max\_sqrt\_num\_cells} of 64 but a higher number for \texttt{num\_leaves} of 100, compared to the default values defined in Section \ref{sec: X-PSI settings} used by the 3C50 run. The effect of these changes on the likelihood value for a given parameter vector is expected to be minimal. However, the variation in value is not consistent for different vectors, and is not necessarily always higher or lower for one of the resolution settings.
Being an earlier dataset, the Delta run incorporated the $\beta$ parameter combining distance and effective area priors, while the 3C50 run keeps the distance fixed. 

The derived marginalized 1-D 68\% credible interval (\CI)\footnote{All results reported in text correspond to the median and the marginalized 1-D 68\% credible interval unless explicitly stated otherwise.} for radius using the Delta dataset is $10.39^{+0.23}_{-0.27}$ km (Figure \ref{appendixfig: ST-U delta all params}), and using the 3C50 dataset is $10.01^{+0.13}_{-0.16}$ km (Figure \ref{fig: ST-U 3C50 MR}), thus overlapping with each other. The posterior shifts towards lower values and gets narrower for the 3C50 run. The shift is likely a consequence of the 3C50 data being cleaner and involving different observing periods. Also considering the differences in resolution settings, and accompanying differences in some of the other inferred parameters, particularly $N_{\rm H}$, the change in inferred radius is not necessarily surprising. As for the tightening of the intervals for the 3C50 run, while it is the opposite of what is typically expected upon a reduction in counts \citep[see, e.g., ][]{Lo13, Psaltis14, Salmi22}, it could be explained as a consequence of a combination of one or many of the following factors -- the 3C50 dataset being cleaner, the distance not being fixed for the Delta run, and/or the decrement in \Multinest resolution for 3C50 simply not having sampled the parameter space sufficiently.

For most model parameters (see Figures \ref{appendixfig: ST-U delta all params} and \ref{appendixfig: ST-U 3C50 bkg params}\footnote{We provide the full set of corner plots in Appendix \ref{appendix: figs}, which include the median and \CI, for all parameters of all models discussed in this letter. Any parameter overlap referred to in text without quoted numbers can be seen in these plots.}), we similarly find an overlap of the \CI. Only certain model parameters and derived quantities such as compactness, neutral hydrogen column density, primary spot colatitude and radius, and secondary spot phase do not. Nevertheless, considering the stringent statistical constraints on these parameters, and the general proximity of their median values (except $N_{\rm H}$), we do not find it to be particularly of concern. The inferred  $N_{\rm H}$ is significantly higher for the 3C50 run which could be explained by the difference in the shape of the spectra between the two datasets. The lower channels of the 3C50 dataset have a particularly low number of counts compared to the Delta dataset, which can be attributed to stronger interstellar attenuation. Both datasets indicate a surface emission pattern consisting of both spots near the north pole with a large primary spot encompassing the pole and a small secondary spot next to it, of slightly lower colatitude and nearly anti-phased.

We perform a graphical posterior predictive check to inspect the model's ability to fit the data. This is done by taking an ensemble of posterior samples and generating (Poisson) standardized residuals over the full phase-channel intervals.  Upon visual inspection of the residuals, we find no distinctly identifiable systematic structures in either case\footnote{See residual generated by \texttt{CST+PDT} model in Section \ref{sec: CST+PDT NICER results} for an example of a clean residual plot.}, suggesting that this model is generally capable of explaining either dataset when the background is unconstrained, and yields results that are consistent with each other.\\

\subsubsection{Effect of NICER background constraints} 

In the absence of any background constraints, for both datasets, the inferred background is very high and constitutes the bulk of the total spectral count rate across all channels compared to the contribution of the spots' emission (see, e.g., left panel of Figure \ref{fig: ST-U 3C50 BKG plot}).

Let us first consider only the cases where we impose a lower limit on the background. For the Delta run involving the space weather estimate and \texttt{elsewhere} component, we find no significant differences for the posterior distributions of any of the model parameters when compared to the run without any background constraint, with the \CI overlapping well with each other for all parameters. The \texttt{elsewhere} component is inferred as having a broad posterior in very low temperatures (${\rm log_{10}} (T_{\rm else}[\rm K]) = 5.32\pm0.20$) which therefore does not have any major contribution in the observed energy range.
The inferred background is still high, although this time, the associated standard deviations are slightly broader below channel 40. The residuals are virtually indistinguishable compared to those of the background unconstrained run.

For the 3C50 dataset, we tested for two scenarios when only imposing a lower background limit: the first involving only the 3C50 instrument background estimate; the second involving an estimate of the AGN contribution in combination with the first. For the instrument-only case, we also test for $n_{l} \in [1,4]$  (see Section \ref{sec:sampling} for definition of $n_{l}$). In all cases, there is good overlap for all parameters,  typically at the \CI level, except the size and location of the primary spot (see Figure \ref{appendixfig: ST-U 3C50 uncertainty params}). For these parameters, the $n_{l}=1$ case slightly differs from the rest, but is still very much in the vicinity, especially accounting for their extremely narrow posteriors (for all $n_{l}$). 

The deviations of the posteriors do not exhibit any unidirectional trend for any of the parameters when going from $n_{l}=1$ to 4. This is likely due to insufficient sampling resolution\footnote{This run used the same settings as the run without background constraints.}, and the strong influence of one or a very few channels when using the raw 3C50 background spectrum, as was found in \citet{Salmi22}. They all infer roughly similar background contributions, barring slight variations in the breadth of the associated standard deviations. The inferred backgrounds also follow the same trend as the run without any constraints, being very high. 

\begin{figure}[t]
    \centering
    \includegraphics[width=0.47\textwidth]{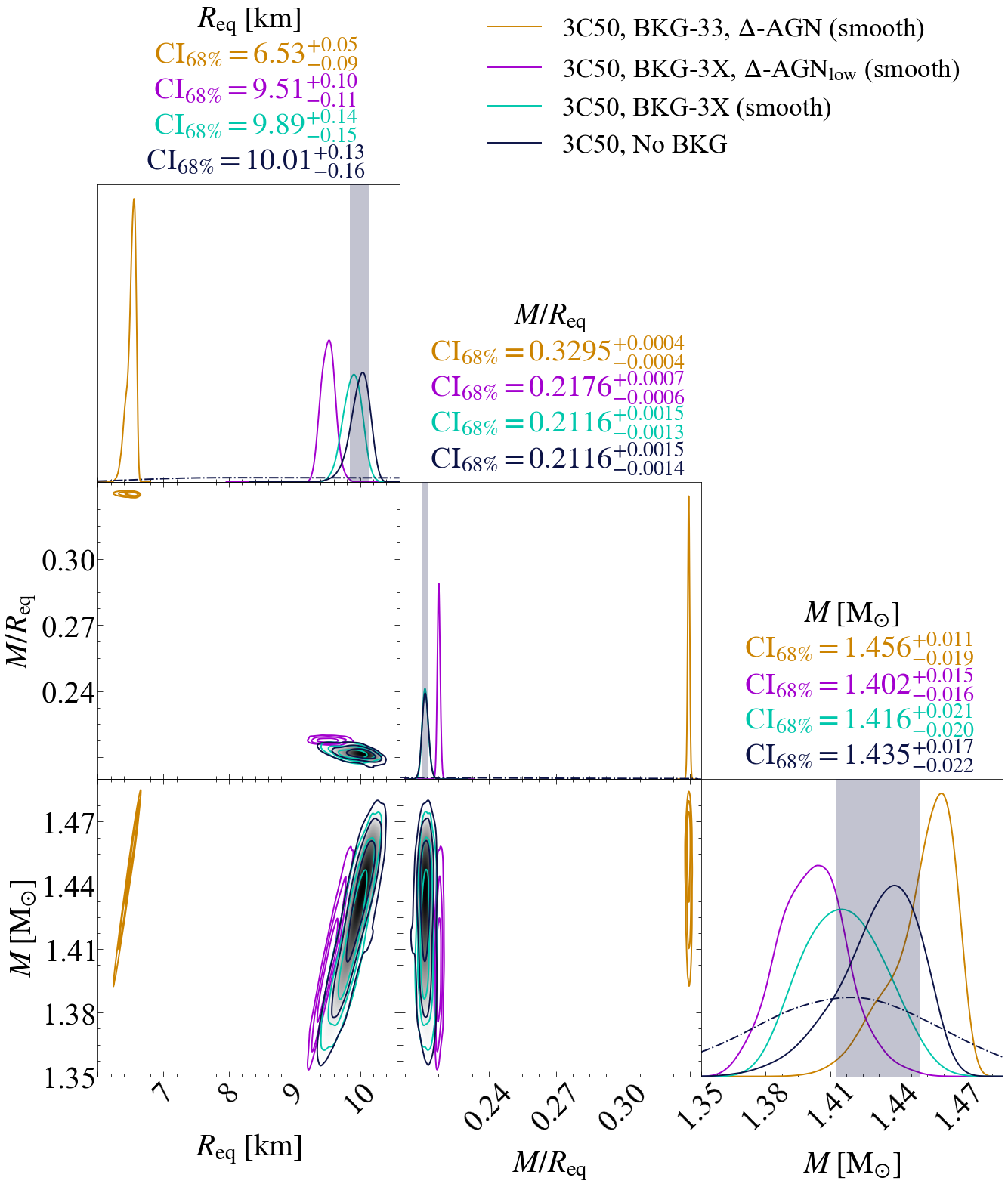}
    \caption{1-D and 2-D marginalized posteriors of mass, radius and compactness inferred by \texttt{ST-U}  for the different implementations of background constraints on the 3C50 dataset. The shaded region in the 1-D marginalized posterior denotes the 68.3\% credible interval for the very last model listed in the legend. The contours in the 2-D marginalized posterior for a given model denote the 68.3\%, 95.4\%, and 99.7\% credible intervals, and the posterior for the last model in the legend is shaded. The dash-dotted lines represent priors for these parameters (also for all other corner plots in this Letter). In the legend, 3C50 refers to the dataset used, BKG refers to the background and the two numbers that follow correspond to $n_{l}$ and $n_{u}$ ($n_{u}$=X indicates an absence of an upper limit), $\Delta$-AGN refers to the inclusion of the Delta AGN spectra (scaled to 3C50 exposure time) when defining lower and upper background limits. The inferences differ vastly upon inclusion of an upper background limit.}
    \label{fig: ST-U 3C50 MR}
\end{figure}

Going forward we choose $n_{l} = 3$ (and for $n_{u}$ too, wherever applicable), for which we also tested the effect of smoothing the background. Similar to \citet{Salmi22}, we used a Savitzky–Golay filter with a window length of 17 (number of coefficients) and a fifth order polynomial. We find that the inferred radius of $9.89^{+0.14}_{-0.15}$ km now matches more closely with that of the background unconstrained case (Figure \ref{fig: ST-U 3C50 MR}), compared to a radius value of $9.47^{+0.07}_{-0.10}$ km for the unsmoothed version. The same trend is observed for almost all model parameters. The residuals are nearly identical and the backgrounds inferred are very similar with slightly broader intervals for the smoothed run. 

In the second scenario, most of the 3C50 testing was done using an AGN spectrum before it was filtered using the 3C50 technique, and therefore involved a higher count rate. The limits on the AGN were implemented as described in Section \ref{subsec: likelihood}. We raise the lower background limit by adding the lower AGN limit onto the 3C50 instrument background (smoothed, $n_{l} = 3$; used hereon). This run again infers a somewhat similar radius of $9.51^{+0.11}_{-0.11}$ km. The slight drop, given the tight constraints for all the runs however, implies that the 68\% credible interval of this run does not overlap with the corresponding interval of the lower instrument-background-only and the free background cases (Figure \ref{fig: ST-U 3C50 MR}). 

\begin{figure*}[t]
    \centering
    \includegraphics[width=\textwidth]{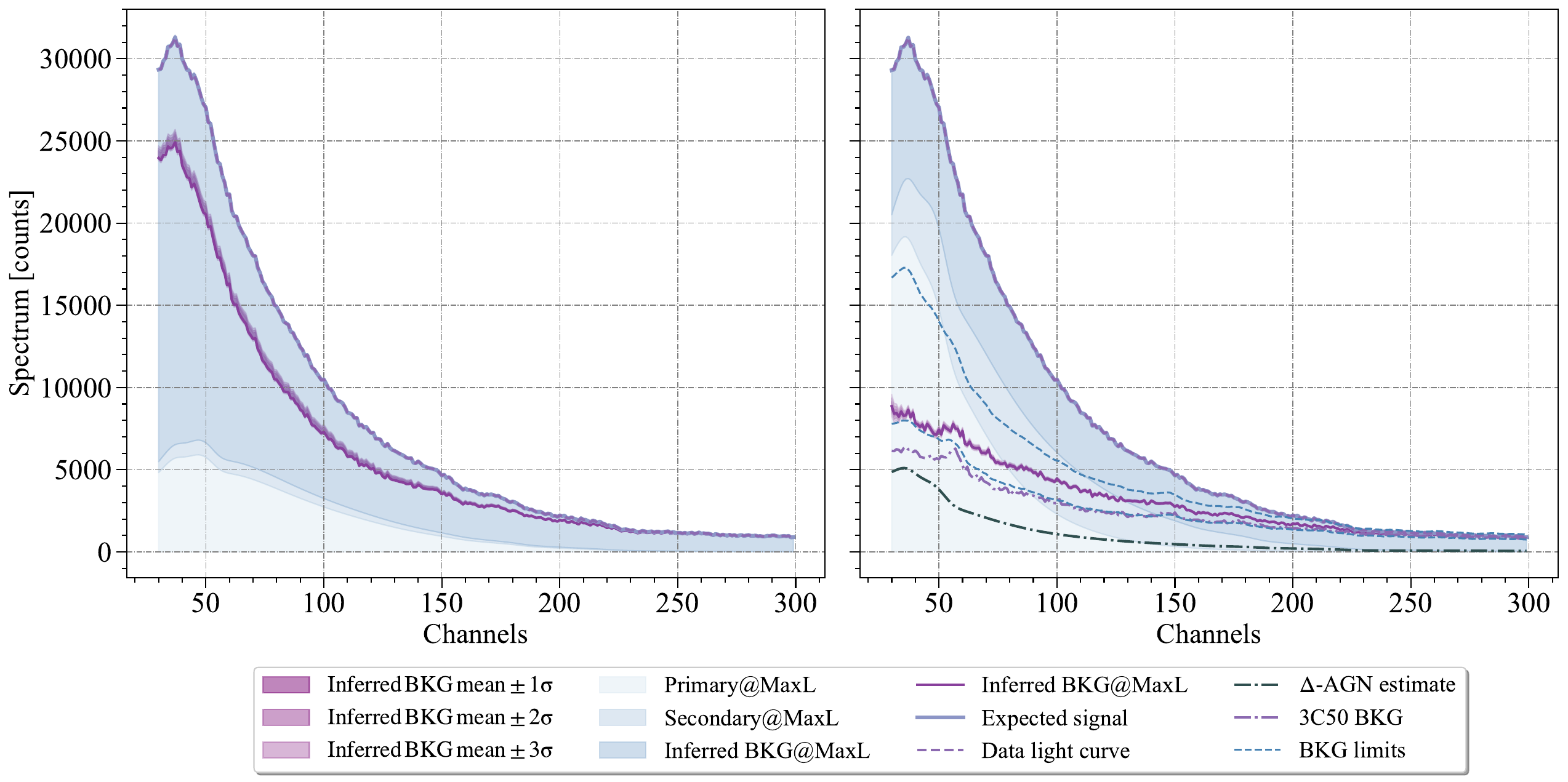}
    \caption{Inferred background and spot contributions by \texttt{ST-U} in the absence of any background constraints (left), and in the presence of both lower and upper background constraints (right) involving the instrument background and the Delta AGN spectrum (scaled to 3C50 exposure time), as applied to the 3C50 dataset. The inferred background constitutes the bulk of the observed data in the absence of an upper background constraint. For more details of the Figure elements and notations, see caption of Figure 5 in \citet{Vinciguerra23b}}
    \label{fig: ST-U 3C50 BKG plot}
\end{figure*}

Spot region parameters aside, the other parameters are in good agreement between the runs involving no background limits or only a lower limit (see Figure \ref{appendixfig: ST-U 3C50 bkg params}). The posterior distributions of most hot region parameters are nearly completely detached between this run and the background-unconstrained run, while some of them show some degree of overlap with the instrument-background-only run. All the parameters can still be said to be very close to each other, again considering the stringent statistical constraints. Again, the residuals and inferred backgrounds are highly similar between them. Overall, the presence of only lower background limits do not seem to affect the results much.

The results change drastically upon the introduction of an upper background limit. The radius drops steeply to $6.53^{+0.05}_{-0.09}$ km, making the star far more compact as can be seen in Figure \ref{fig: ST-U 3C50 MR}. We also see a significant increase in the value of $N_{\rm H}$ and a slight decrease in the value of $\alpha_{\rm XTI}$, compared to the other runs discussed so far, all of which contribute to a lower number of counts. The inferred geometries are also very different.

\begin{figure}[t]
    \centering
    \includegraphics[width=0.47\textwidth]{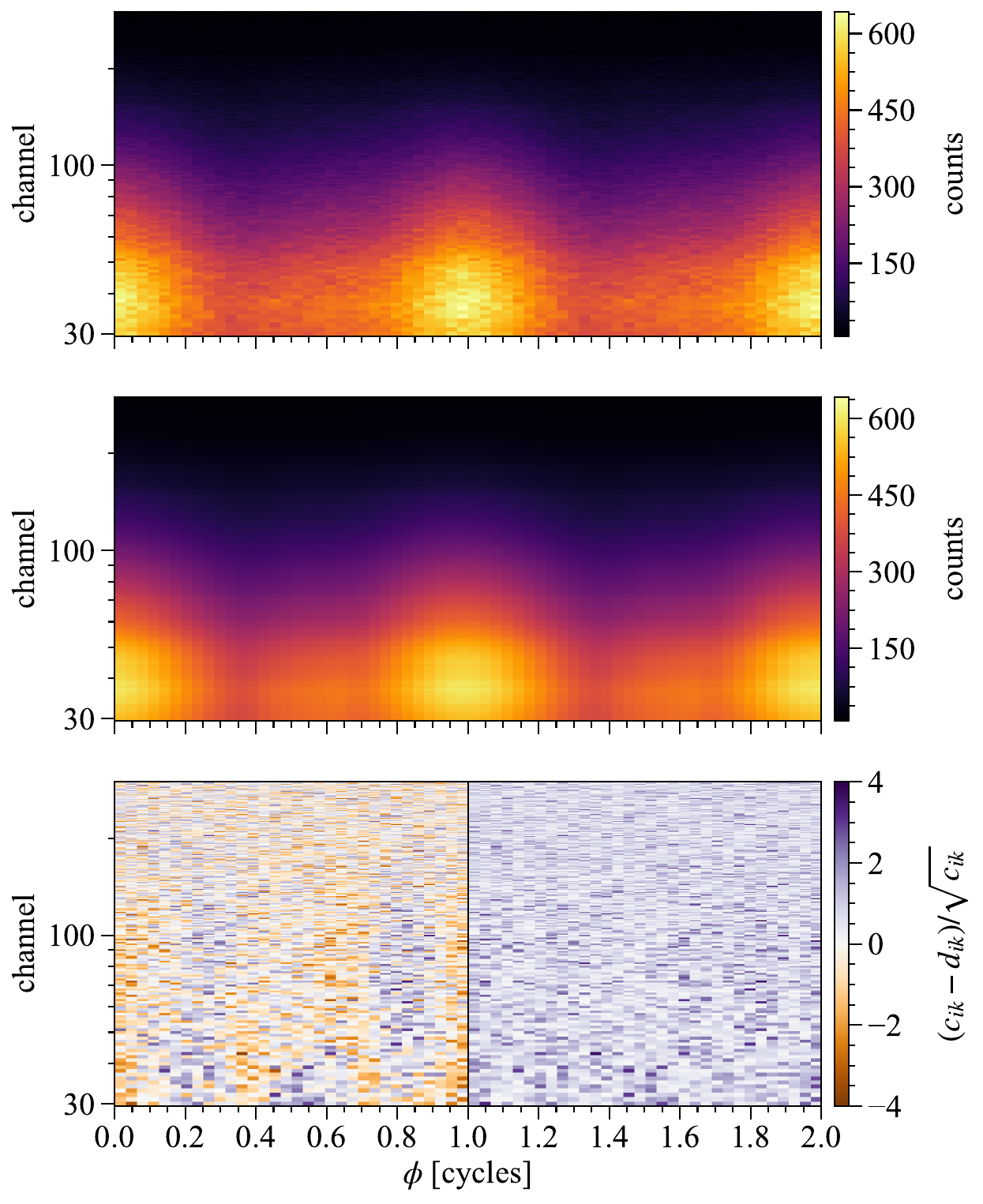}
    \caption{Top panel: 3C50 data. Middle panel: Posterior expected count numbers generated by the \texttt{ST-U} model when imposing both lower and upper background limits. Bottom panel: Poisson standardized residuals ($c_{ik}$ and $d_{ik}$ represent the model-generated and data counts resolved over phases $i$ and channels $k$). Prominently visible systematic structures indicate poor model performance.}
    \label{fig: ST-U 3C50BKG_Agn_smooth_residuals}
\end{figure}

The inferred background is now bound to be within the specified limits and we find that it is particularly low in the channels $\lesssim 50$, where it is close to the lower limits (Figure \ref{fig: ST-U 3C50 BKG plot}). Looking at the residuals for this run (Figure \ref{fig: ST-U 3C50BKG_Agn_smooth_residuals}), we find prominent phase-dependent structures where the model alternatively either over-predicts or under-predicts the data. This indicates a clear model deficiency for \texttt{ST-U} in explaining the data in the presence of an upper background constraint.

\subsubsection{Effect of XMM-Newton constraints}

We attempted joint fits for NICER and XMM-Newton only on the Delta dataset for the \texttt{ST-U} model. The addition of XMM-Newton imposes lower and upper background constraints on our models, as explained in Section \ref{subsec: likelihood}. We do not add any separate NICER background constraints for this run. We find that this has similar effects to the imposition of an upper NICER background limit. We infer a low radius value of $6.83\pm0.07$ km resulting in a highly compact star. The effective areas for both telescope instruments drop to unrealistic values, requiring a scaling factor of $0.55^{+0.008}_{-0.007}$ for NICER and $0.38^{+0.007}_{-0.006}$ for XMM-Newton. The residuals also show similar phase-dependent residual structures but even more pronounced this time around, once again showing that \texttt{ST-U} is unable to explain the data in the presence of upper background constraints.

\subsection{ST+PDT}

\subsubsection{Effect of datasets}

Since \texttt{ST-U} is our simplest model and is clearly unable to explain data in the presence of upper background constraints, we also check for model consistencies between the datasets using \texttt{ST+PDT}, one of our more capable models in this regard (as demonstrated in the coming paragraphs). 
\texttt{ST+PDT} appeared to be promising based on tests on earlier datasets.
We directly tested the Delta dataset inclusive of space weather estimates on this model. The closest matching run with the 3C50 dataset is one consisting of a lower instrument-background-only constraint. 

Datasets and background limits aside, the two runs also differ in their resolution settings. The Delta run uses higher than default \Multinest resolution settings of LP 8000, SE 0.1, motivated by tests on an initial dataset. The 3C50 runs on the other hand use all of the default settings except for a lower LP of 2000 to save on computational resources for testing purposes. The Delta run also uses slightly higher \ac{X-PSI} resolution for \texttt{min\_sqrt\_num\_cells} of 16 and \texttt{num\_leaves} of 100 whereas the 3C50 runs stick to the default values. 

The Delta run infers a radius of $11.14^{+0.46}_{-0.40}$ km (Figure \ref{appendixfig: ST+PDT Delta params}) while the corresponding 3C50 run infers a value of $10.33^{+0.41}_{-0.39}$ km (Figure \ref{fig: ST+PDT 3C50 MR}), thus overlapping marginally with each other. This trend of reduction in radius for the 3C50 dataset is therefore somewhat consistent with what was observed for \texttt{ST-U}, but more pronounced. Unlike \texttt{ST-U}, the $N_{\rm H}$ values are very similar for the two runs. The differences in radii can therefore be attributed to a combination of the variations in the other parameters, the resolution differences and the inherent differences between the two datasets. 
 
The spot geometry parameters are generally similar to each other between the two runs (see Figures \ref{appendixfig: ST+PDT Delta params} and \ref{appendixfig: ST+PDT 3C50 params}), but are quite different from the patterns inferred by \texttt{ST-U}\footnote{All models attempted consisting of only single temperature regions infer roughly similar geometries to \texttt{ST-U}.}. This model infers a  large \texttt{ST} spot encompassing the north pole and a smaller \texttt{PDT} spot near the equator. Both Delta and 3C50 runs infer overlapping \CI for the \texttt{ST} angular size and temperature. The \texttt{ST} colatitude credible intervals overlap at a 95\% level. For the Delta run, we find two adjacent but disjoint posterior modes for the phase of the \texttt{ST} spot. The primary mode contains a much larger posterior mass such that the secondary mode only shows up in the 95\% credible interval. The secondary mode overlaps well with the inferred values of the 3C50 run.

The \texttt{PDT} spot consists of a very small and high temperature \textit{superseding} component compared to the larger and low temperature \textit{ceding} component. For the Delta run, all hot spot parameters (except the \textit{ceding} component's colatitude) also exhibit two posterior modes, which are close to each other. Similar to the \texttt{ST} phase, one of the modes has a much larger posterior mass than the other. Between the two modes, some of the parameters share the 99\% credible interval while it is disjoint for others. The secondary modes for all these parameters overlap significantly with the corresponding inferred values of the 3C50 run. The \textit{ceding} component's colatitude 68\% credible intervals also overlap for the two runs. 

The secondary modes of all the parameters, wherever present, extend nominally towards lower radii. This could indicate that the dominant primary mode of the Delta run is either not identified by the 3C50 run given its lower resolution settings, or that the cleaner 3C50 dataset does not favor such solutions. We find the latter to be more likely given the similar posterior widths of the two modes and their close proximity to each other, which \Multinest should therefore have identified, especially considering the larger posterior mass of the primary mode recognized for the Delta run.

Both runs infer similarly high background count rates as was seen for \texttt{ST-U} in the absence of any upper constraint. Neither run shows any distinct residual structure suggesting that model output resembles the data well. The results obtained using either dataset can also be considered to be generally in agreement with each other.

\subsubsection{Effect of NICER background constraints}

The \texttt{ST+PDT} model was tested on the 3C50 dataset for three cases: without any background constraints, with only lower instrument background constraints, and with both lower and upper constraints involving the 3C50 AGN estimate. Note that here we are utilizing the actual 3C50 AGN spectrum instead of scaling the Delta AGN spectrum to 3C50 exposure times as was done during the inference runs using \texttt{ST-U}. 

\begin{figure}[t]
    \centering
    \includegraphics[width=0.47\textwidth]{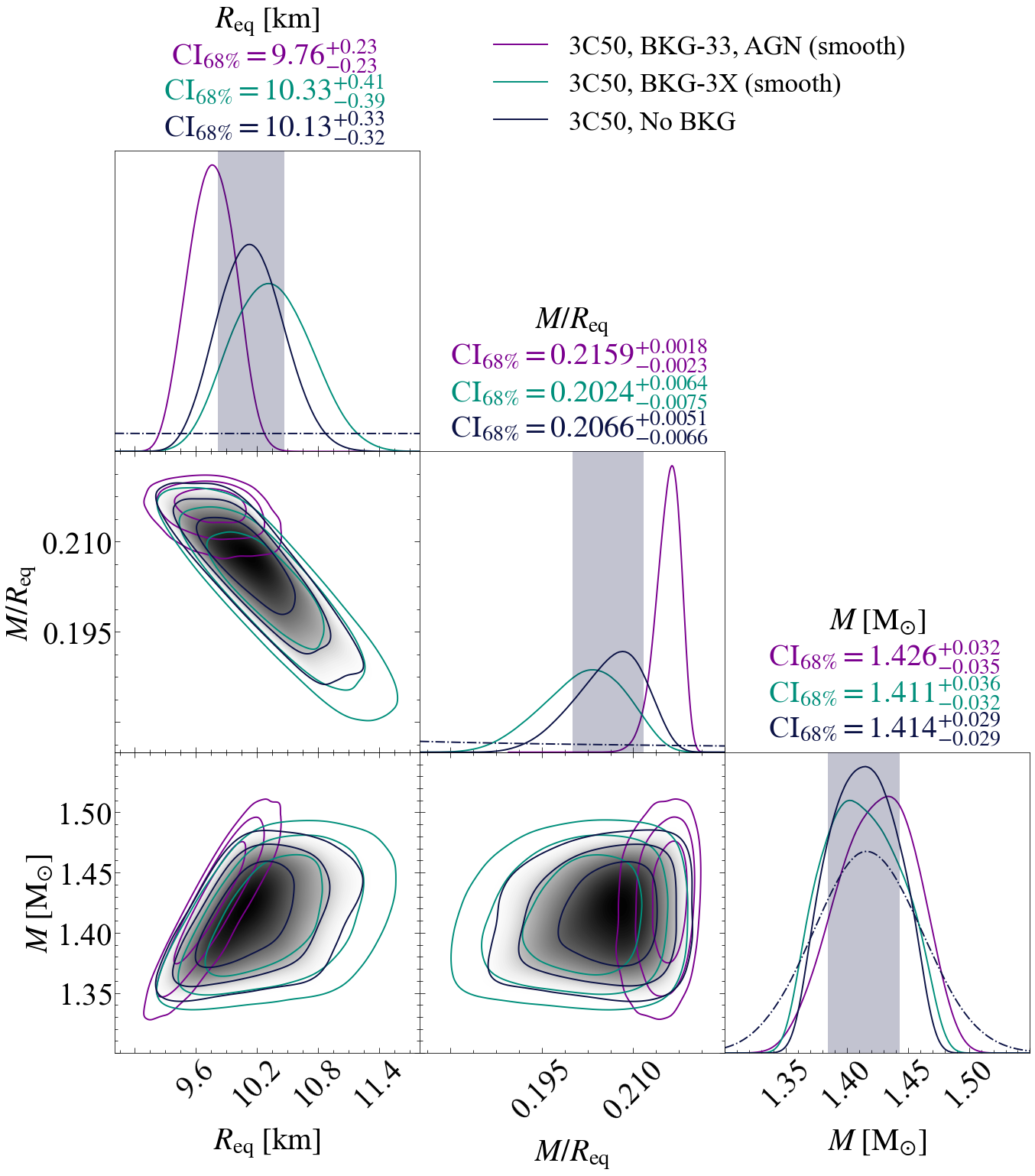}
    \caption{Mass, radius and compactness posteriors inferred by the \texttt{ST+PDT} model configuration for the different implementations of background constraints on the 3C50 dataset. The inclusion of an upper background limit tightens the radius posterior and favors a more compact solution.}
    \label{fig: ST+PDT 3C50 MR}
\end{figure}

The radius posteriors for all three runs overlap well with each other, as shown in Figure \ref{fig: ST+PDT 3C50 MR}. We obtain a radius of $10.13^{+0.33}_{-0.32}$ km for the free background run, and $9.76\pm 0.23$ km when we impose both limits. The posterior is therefore much tighter in the presence of an upper background limit, with a preference for more compact solutions. 

The other non-spot parameters are also consistent between the three runs, particularly for the runs without upper background limits (Figure \ref{appendixfig: ST+PDT 3C50 params}). The run with upper limits infers slightly higher values for $\alpha_{\rm XTI}$ and lower values for $N_{\rm H}$ which could partially account for the lower radii.
The main differences for the upper background constrained runs show up in the spot geometries. In the presence of upper background limits, the \texttt{ST} is inferred to be smaller than the \texttt{PDT} and is quite offset from the north pole. The \texttt{PDT}, instead of being near the equator, now encompasses the south pole, thus contributing significantly to the unpulsed emission. The only similarity between this run and ones without upper background limits is that the \texttt{PDT} \textit{ceding} component is much larger with respect to the \textit{superseding} component. The \textit{ceding} component inferred is larger than that inferred by the other two runs. The \textit{superseding} component's angular radius contours, although not consistent with the other two runs, can be said to infer similar values given the narrow posteriors. As for the spot temperatures, the upper background constrained run infers a much hotter \texttt{ST} region and a slightly cooler \texttt{PDT} \textit{superseding} component. The \texttt{PDT} \textit{ceding} component temperature is similar between all three runs.

\begin{figure*}[t]
    \centering
    \includegraphics[width=\textwidth]{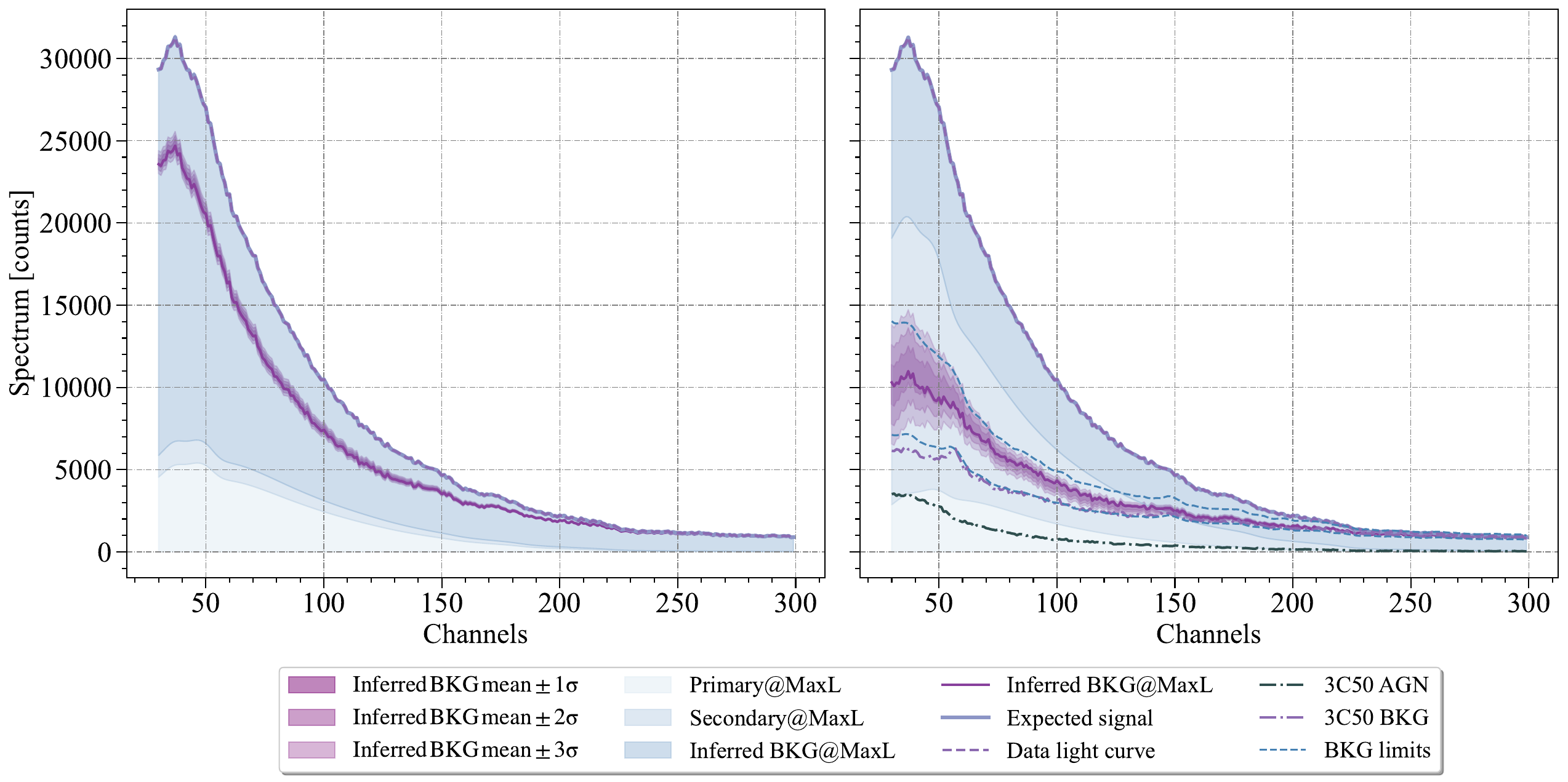}
    \caption{Inferred background and spot contributions by \texttt{ST+PDT} in the absence of background constraints (left), and in the presence of both lower and upper 3C50 background limits (right) involving the instrument background and the 3C50 AGN spectrum, as applied to the 3C50 dataset. Similarly to ST-U, this model also infers a very high background contribution in the absence of upper limits.}
    \label{fig: ST+PDT 3C50 bkg}
\end{figure*}

The inferred background for the runs without upper constraints are largely similar. With the imposition of the upper limit, the inferred background seems to sit about halfway between the two limits as seen in Figure \ref{fig: ST+PDT 3C50 bkg}. The residuals have no apparent features present that would indicate poor model performance. 

\section{Production analyses -- CST+PDT}
\label{sec: Prod analyses}

Having performed our exploratory analyses, we arrive at \texttt{CST+PDT} as our final model in this work, based on model evidence comparisons (as explained later in this section). For the headline result version of this model, we provide all the inferred posterior and maximum likelihood values of the model parameters in Table \ref{tab: CST+PDT}. Having established the consistency between datasets, we only analyze the 3C50 dataset for this model because it is cleaner, and allows us to place both lower and upper background constraints. 

For this model, in addition to default \Multinest resolution runs, we also conducted high resolution runs (simply referred to as high-res runs hereon) using LP 20,000 albeit having reduced certain \ac{X-PSI} resolution settings (\texttt{sqrt\_num\_cells} = 18, \texttt{max\_sqrt\_num\_cells} = 32, \texttt{num\_leaves} = 32, \texttt{num\_energies} = 64). This choice enables us to more thoroughly explore the parameter space while balancing out computational expense\footnote{While this choice affects the likelihood calculations, the deviations are at a level that is not expected to have a major effect on the overall shape of the posterior modes for such a model based on the tests conducted by \citet{Choudhury24_raytracing}.}.  For these high-res runs exclusively, we utilized the latest PPTA-DR3 priors on the mass, inclination, and distance. \\

\subsection{Effect of NICER background constraints}
\label{sec: CST+PDT NICER results}

The high-res runs were conducted for two cases only, namely, with no background constraint and with both lower and upper background constraints using the 3C50 AGN estimate. We have also tested the application of only a lower instrument background constraint for the default resolution runs. 

\begin{figure}[t]
    \centering
    \includegraphics[width=0.47\textwidth]{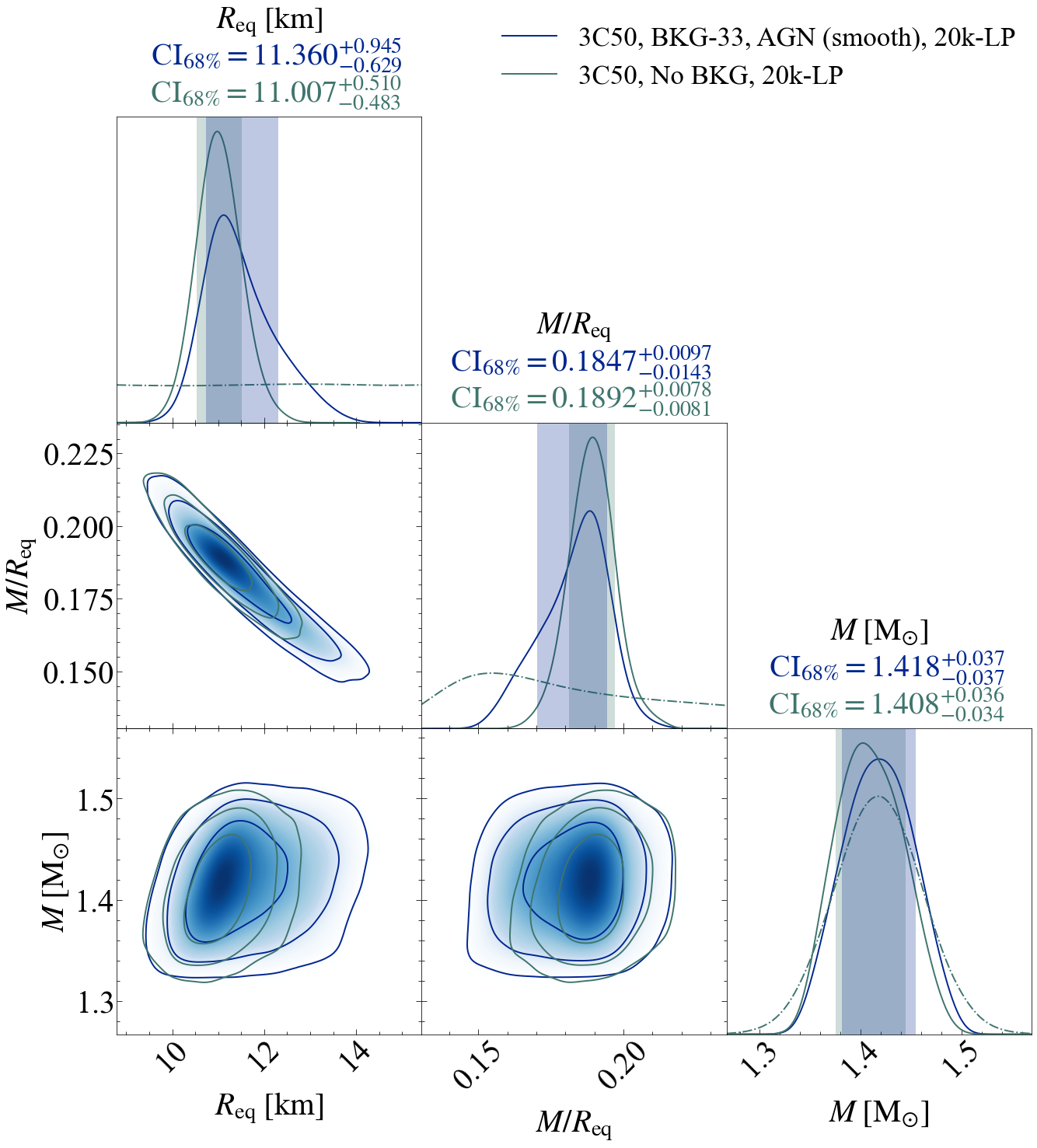}
    \caption{Mass, radius and compactness posteriors inferred by the high \Multinest resolution \texttt{CST+PDT} runs, both in the absence of any background constraints, and in the presence of lower and upper background constraints involving the instrument background and 3C50 AGN spectrum. The 2-D marginalized posteriors of the latter model, that constitutes our headline result, are shaded in blue. Both runs infer highly consistent radius posteriors, with the background-constrained run consisting of a distribution tail that extends slightly more towards higher radii.}
    \label{fig: CST+PDT 3c50 MR}
\end{figure}

In Figure \ref{fig: CST+PDT 3c50 MR}, we show that for the high-res runs, in the absence of any background constraints, we infer a radius of $11.01^{+0.51}_{-0.48}$ km; and with both lower and upper limits in place, \textbf{we infer a radius of $\mathbf {11.36^{+0.95}_{-0.63}}$ km}, our headline result for this letter. The two radius intervals are therefore well in agreement with each other, with the latter consisting of a slightly higher median and a longer tail towards higher radii, but otherwise encompassing the former. The narrower posterior for the run without any background information could be attributed to the presence of solutions with much higher likelihood values whose inferred backgrounds are far higher than those allowed by our upper background constraint. 

The default resolution run with only a lower limit inferred a radius of $10.88\pm0.56$ km, thus overlapping with the aforementioned values. The lower median and the overall posterior being shifted towards lower radii is likely due to the use of PPTA-DR2.5 priors for this run, along with some minor effects of using different resolution settings. We deduce this based on the high degree of overlap between this run and the background-unconstrained default resolution PPTA-DR2.5 run, for which we infer a radius of $10.91 \pm 0.58$ km. If using lower \Multinest resolution settings were the main reason for the lower radius inferred, we would expect stochastic processes to vary the results more between these runs. The differing X-PSI resolution settings are expected to have only a minor influence on the inferred radius. 

Since \texttt{CST+PDT} encompasses the \texttt{ST+PDT} model, we also compare the inferences between these two models to see how addition of a \textit{masking} component to the primary spot affects our results. The radius inferred by the high-res run using no background constraints does not overlap with the corresponding run of \texttt{ST+PDT} at a \CI level. However, given their general proximity, the differences in resolution between these runs, the different radio priors used, and the observed trend of moving towards higher radii for the PPTA-DR3 radio priors, these solutions can be considered to be in agreement with each other. The same cannot be argued, however, when we compare the corresponding cases using both lower and upper background constraints due to the much larger differences between the inferred radii, and the very narrow radius posterior inferred by \texttt{ST+PDT}. 

\begin{figure*}[t]
    \centering
    \includegraphics[width=0.8\textwidth]{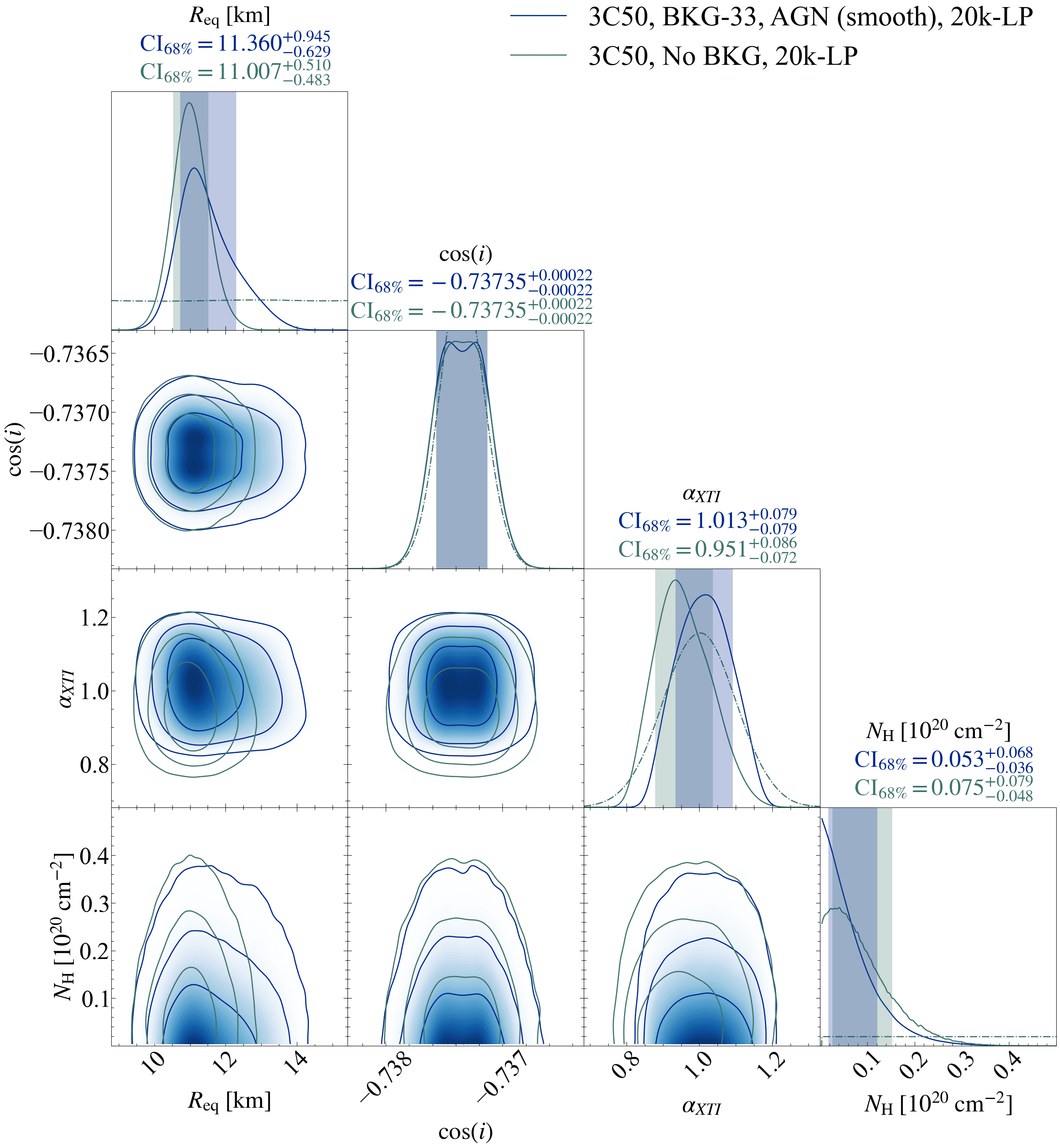}
    \caption{Radius, inclination, instrument scaling and $N_{\rm H}$ posteriors inferred by the high \Multinest resolution \texttt{CST+PDT} runs for the two different implementations of background constraints on the 3C50 dataset.}
    \label{fig: CST+PDT 3c50 other}
\end{figure*}

The other non-spot parameters (Figure \ref{fig: CST+PDT 3c50 other}) overlap well between the two high-res runs at the \CI level\footnote{We abstain from further describing the results of the run with only a lower background limit, due to its use of the older radio priors, default resolution settings, and its overall overlap with the corresponding run with no background constraints. For more details on the results of the default resolution runs, Figure \ref{appendixfig: CST+PDT params}}. The run with both background limits in place infers slightly higher values for $\alpha_{\rm XTI}$ and lower values for $N_{\rm H}$. For \texttt{ST+PDT}, similar trends seen for these two parameters partially explained the lower radii values inferred compared to the corresponding background-unconstrained case. For this model however, we see an increment in the inferred radii values instead, due to the difference in the \texttt{PDT} spot geometry (described in the coming paragraphs). For \texttt{CST+PDT}, the differences in $\alpha_{\rm XTI}$, $N_{\rm H}$ and radii, compared to the background-unconstrained case, all compensate for the missing unpulsed emission upon imposition of an upper background limit.

\begin{figure*}
    \centering
    \includegraphics[width=1.0\textwidth]{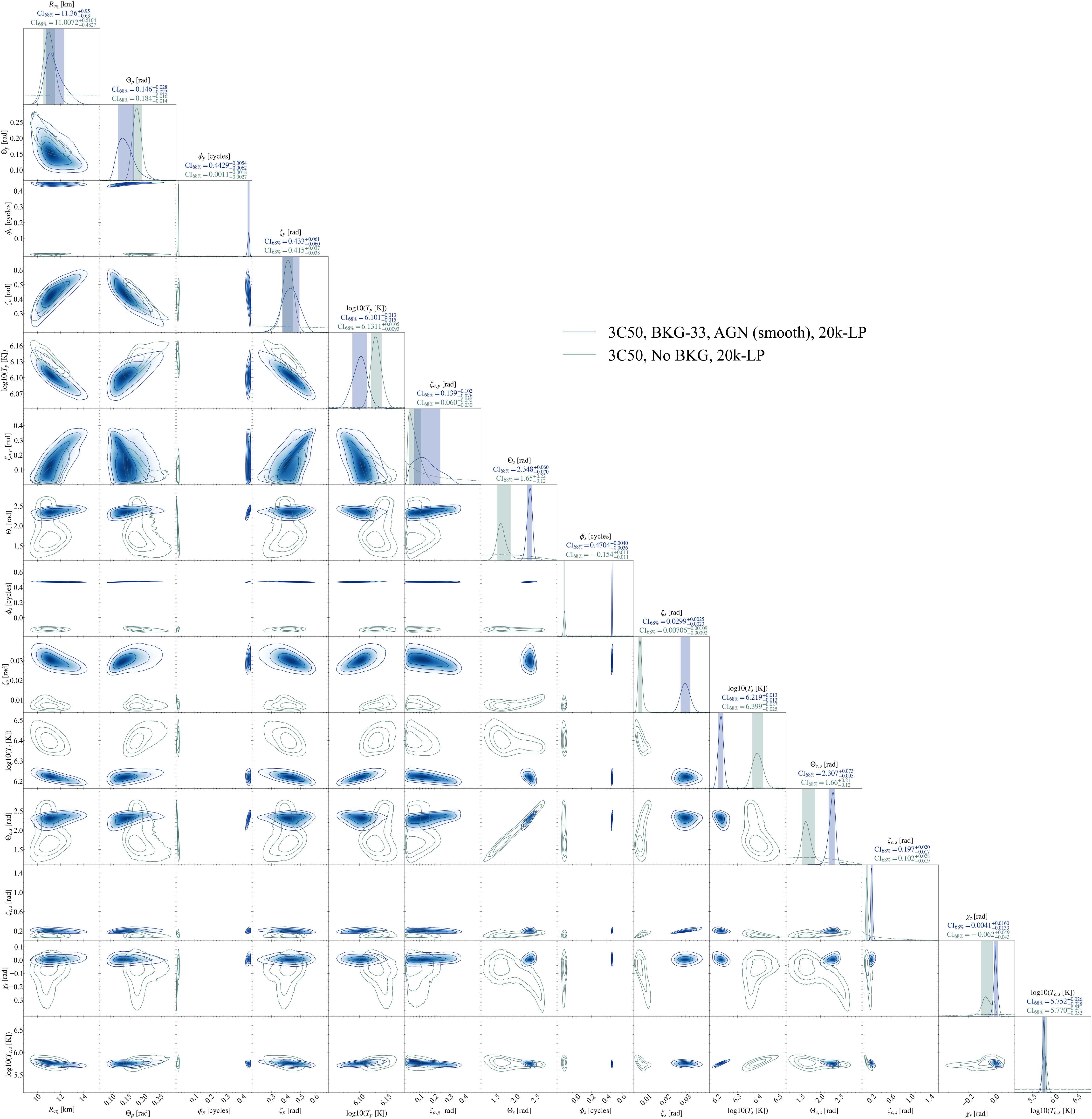}
    \caption{Radius and spot geometry posteriors inferred by the high \Multinest resolution CST+PDT runs for the two different implementations of background constraints on the 3C50 dataset.}
    \label{fig: CST+PDT 3c50 geom}
\end{figure*}

\begin{figure*}[t]
    \centering
    \includegraphics[width=1.0\textwidth]{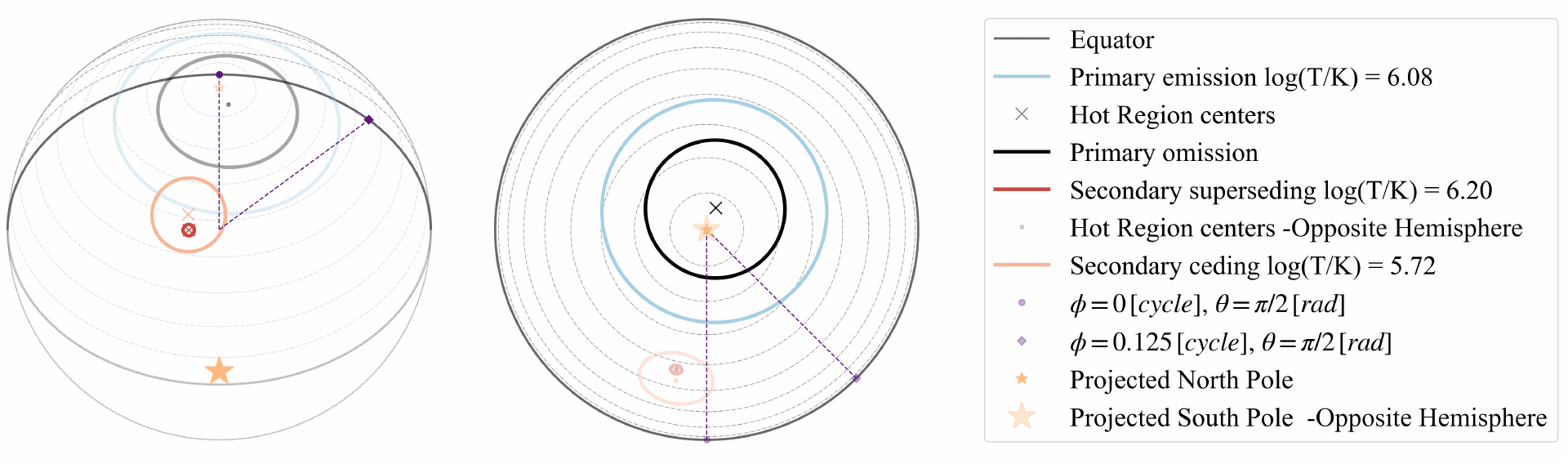}
    \caption{Projections of the inferred maximum likelihood (ML) geometry by the high \Multinest resolution CST+PDT run involving both lower and upper background constraints as viewed from the observer inclination (left) and from the north pole (right). Hot spots in the opposite hemispheres are indicated using dimmed colors. Lensing effects are not shown in these projections. The ML geometry consists of a ring encompassing the north pole, and a two-temperature spot in the southern hemisphere.}
    \label{fig: CST+PDT projection}
\end{figure*}

The main differences show up in some of the hot region parameters (Figure \ref{fig: CST+PDT 3c50 geom}). The \texttt{CST} hot regions are still somewhat similar between the two runs. The angular radius of the emitting component is about the same in both cases. The size of the \textit{masking} component inferred, however, is different. When using no background constraints, the posterior peaks at very small angular radii for the \textit{masking} component followed by a gentle drop-off, resulting in a \CI of $0.06^{+0.05}_{-0.03}$ radians. The maximum likelihood solution has a value of 0.037 radians, making the \texttt{CST} spot somewhat degenerate with a \texttt{ST} spot. In the presence of both background limits, a larger hole is preferred, with a \CI of $0.14^{+0.10}_{-0.08}$ radians for the \textit{masking} component. Both \texttt{CST}s have similar locations where they encompass the north pole (see Figure \ref{fig: CST+PDT projection} showing the projection of the spot geometry for the maximum likelihood vector of the high-res run consisting of both background limits). They also have comparable temperatures, even though they do not overlap at the \CI level given their stringent statistical constraints.

The \texttt{PDT} hot regions are comparatively more dissimilar between the two high-res runs, particularly in terms of their location. For the background-unconstrained run, the colatitude posteriors of both the \texttt{PDT} components consist of a main peak (comprising the \CI), that is just south of the equator, and a plateau-like extension towards higher colatitudes. However for the run with both background limits, both \texttt{PDT} components are located half-way between the equator and the south pole, almost at the observer inclination angle, with a tighter posterior mode for their colatitudes. This mode is encompassed within the plateau-like extension of the background-unconstrained run. The \texttt{PDT} regions are also well-separated in phase between the two runs, with a difference of 0.38 cycles between the medians of the \textit{superseding} component's phases, and consisting of very narrow posteriors in each case.

The similarities lie in the pattern of sizes and temperatures of the \texttt{PDT} components between the two runs. In both cases, we infer a very small and high temperature \textit{superseding} component, and a comparatively larger and cooler \textit{ceding} component, similar to the findings of the \texttt{ST+PDT} run without an upper background constraint. For both the \texttt{PDT} components, the run with lower and upper background limits infers slightly larger angular radii such that the \CI do not overlap with the background-unconstrained inferred values. The former also infers a slightly hotter \textit{superseding} component, again with no \CI overlap. The two runs only strictly agree on the temperature of the \textit{ceding} component. 

\begin{figure*}[t]
    \centering
    \includegraphics[width=\textwidth]{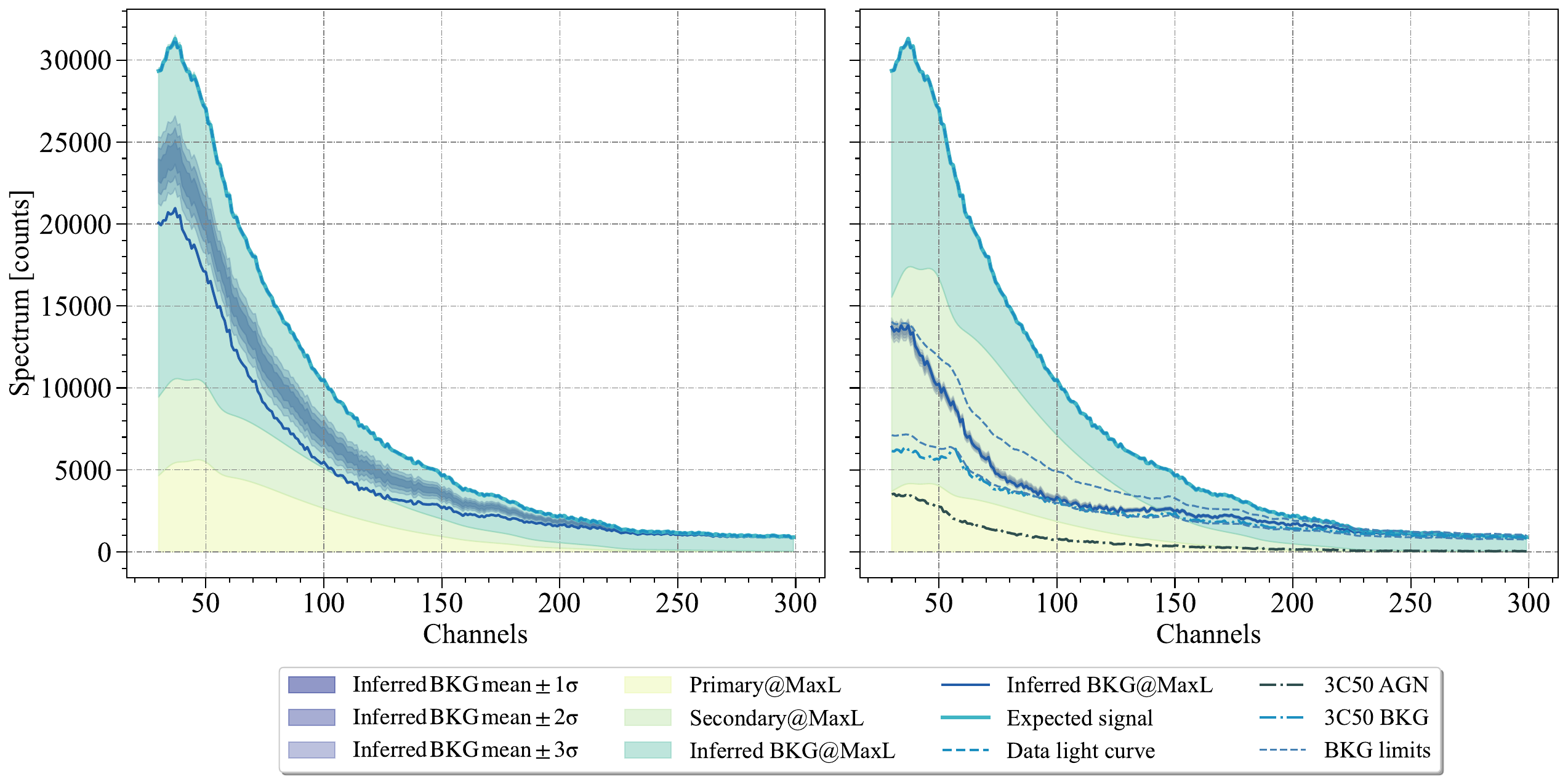}
    \caption{Inferred background and spot contributions by the high \Multinest \texttt{CST+PDT} runs in the absence of background constraints (left), and in the presence of both lower and upper 3C50 background limits (right) involving the instrument background and the 3C50 AGN spectrum, as applied to the 3C50 dataset. Similar to the exploratory models, we once again infer a very high background contribution in the absence of upper limits.}
    \label{fig: CST+PDT 3c50 bkg}
\end{figure*}

Similar to \texttt{ST-U} and \texttt{ST+PDT}, the background-unconstrained run infers a very high background contribution, far beyond the limits we consider, particularly in the lower energy channels (Figure \ref{fig: CST+PDT 3c50 bkg}). Upon imposition of the lower and upper constraints, the inferred background is rather high (within the bounds) for some of the lowest channels ($\sim 30-40$) where it pushes against the upper limit, and then slowly drops and gets rather low for the channel range $\sim 80-150$, pushing against the lower limit. The background-constrained model compensates for the lower overall inferred background (compared to the unconstrained case) by inferring the position of the PDT spot such that it is never eclipsed, which contributes significantly to the unpulsed emission, and lowers the pulse amplitude. 

\begin{figure*}
    \centering
    \includegraphics[width=0.7\textwidth]{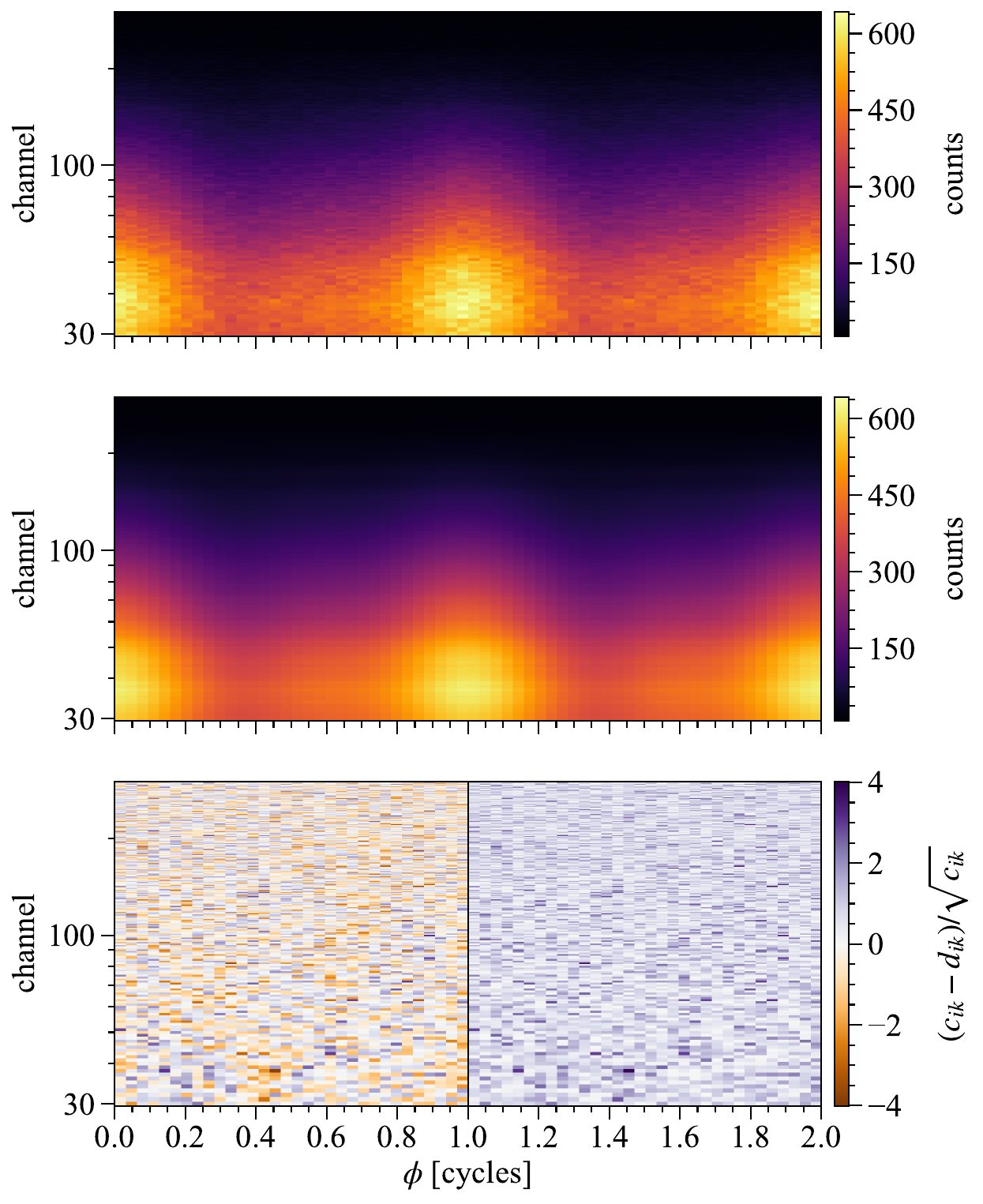}
    \caption{Top panel: 3C50 data. Middle panel: Posterior expected count numbers generated by the high resolution \Multinest \texttt{CST+PDT} run when imposing both lower and upper background limits. Bottom panel: Poisson standardized residuals showing no apparent systematic structures, indicating that the model is capable of explaining the data well.}
    \label{fig: CST+PDT 3c50 residuals}
\end{figure*}

The residuals show no distinctly identifiable systematic structures for either high-res run (Figure \ref{fig: CST+PDT 3c50 residuals}), indicating the model is generally capable of explaining the data, both in the presence and absence of background constraints.

In order to gauge improvement in model performance with respect to the corresponding runs of \texttt{ST+PDT}, we first compare the default resolution \texttt{CST+PDT} runs against \texttt{ST+PDT}, both of which use PPTA-DR2.5 radio priors, to strictly isolate the performance gain by the model alone. Then we compare the default resolution \texttt{CST+PDT} run against the high-res runs to understand the difference in performance in light of the latest PPTA-DR3 radio priors\footnote{The \ac{X-PSI} resolution settings also affect the likelihood, and consequently the evidence calculation too. While it is not possible to disentangle the effects of updating the priors from the effect of resolution changes, the latter is expected to only nominally influence the evidence calculation.}.  

When imposing no background constraints, the default resolution \texttt{CST+PDT} run has a higher log-evidence\footnote{All log-evidence values reported in the letter are in natural log units.} than the \texttt{ST+PDT} run by 3.889 units, which is not a very large difference based on the scale provided by \citet{Kass95}. However, given the larger model space of \texttt{CST+PDT} which encompasses \texttt{ST+PDT}, we select the former as our headline model. The high-res run only nominally improves the log-evidence value by another 0.257 units. Similar trends are also observed for the runs with both lower and upper background limits in place. The default resolution \texttt{CST+PDT} run is higher by 1.395 units of log-evidence compared to the \texttt{ST+PDT} runs, and the high-res run improves the log-evidence by a further 1.230 units.

\subsection{Effect of XMM-Newton constraints}
\label{sec: CST+PDT XMM results}

We also tested the \texttt{CST+PDT} joint fits using NICER and XMM-Newton. We place both upper and lower background constraints on the NICER data, with XMM-Newton providing effectively an additional constraint on the background. This run was performed using the default resolution settings and the PPTA-DR2.5 radio priors. The computational expense of a high resolution joint NICER/XMM-Newton run was estimated to be prohibitive given available resources; taking this into account and the additional cross-calibration uncertainty introduced by the joint fit, we treat this lower resolution run only as a cross-check (particularly since our high resolution NICER-only headline result run already includes background constraints).

\begin{figure}[t]
    \centering
    \includegraphics[width=0.47\textwidth]{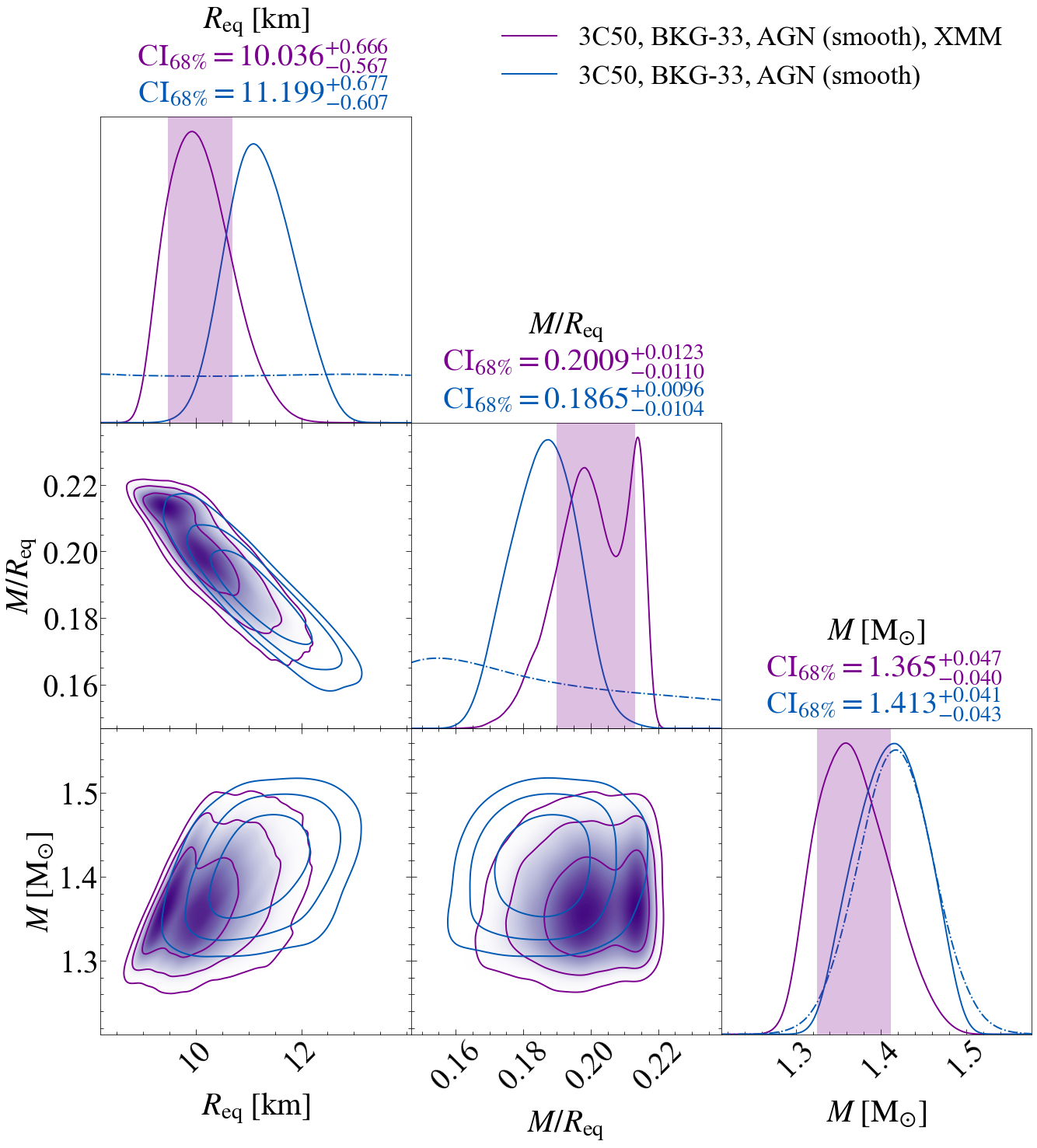}
    \caption{Mass, radius and compactness posteriors inferred by the \texttt{CST+PDT} for the NICER-only run (default resolution) involving both lower and upper background limits, and for the joint NICER-XMM run. The joint NICER-XMM run exhibits a bimodal structure, not seen in any of the previous NICER-only analyses. The less compact mode agrees more with the background-constrained NICER-only analysis for this model.}
    \label{fig: CST+PDT XMM}
\end{figure}

In Figure \ref{fig: CST+PDT XMM}, we show the inferred mass, radius, and compactness for this run, compared against the default resolution NICER-only run with both lower and upper background constraints. The NICER+XMM run infers a radius \CI of $10.04^{+0.67}_{-0.57}$ km, which overlaps with the results of the default resolution NICER-only run including both background limits, but does not with the results of the corresponding high-res run. The median also shifts to a lower value with respect to the NICER-only run for this model. Given the observed trend of moving towards higher radii upon using the latest PPTA-DR3 radio priors, we expect a similar shift upwards for this run, especially considering the drop in the inferred mass, with the median of the mass lying below the lower 1-$\sigma$ limit of the new mass prior. The application of the latest prior thus ought to reduce the offset with the high-res NICER-only run with both background limits in place.

The compactness posterior exhibits a slightly bimodal structure although such modality does not appear separately in the marginalized 1D mass and radius posteriors. The less compact mode overlaps better with the posterior of the NICER-only run. The correlation between radius and compactness results in this more compact mode dragging the radius posterior towards lower values.

The other non-spot parameters (Figure \ref{appendixfig: CST+PDT XMM params}) have very good agreement between the two runs. We obtain a much tighter constraint on the NICER instrument scaling factor with a \CI of $1.04^{+0.02}_{-0.03}$, and also a very tight constraint on the XMM-Newton instrument scaling factor with a \CI of $0.99^{+0.01}_{-0.01}$. The spot geometries are somewhat similar between this run and the NICER-only run with both background limits in place. They more closely resemble each other compared to the runs involving no upper background constraint. 

Once again, we infer a \texttt{CST} region encompassing the north pole, although this time the center is slightly more offset from the pole compared the NICER-only runs. The colatitude credible intervals only overlap at the 95\% level with the default resolution NICER-only run with both background limits. The angular radius of the emitting components agree at a \CI level between these two runs, with the XMM run slightly smaller angular radii. The \CI of the angular radius of the \textit{masking} component strongly overlap with each other, with the XMM run posterior peaking at lower values. The temperatures inferred for the \texttt{CST} region are almost identical.

The colatitude and angular radius of the \texttt{PDT} spot's \textit{superseding} component are nearly identical between the two runs. The corresponding temperatures agree at a \CI level, with the XMM run preferring slightly higher values. The credible intervals of their phases only overlap at the 95\% level, but can be considered to be roughly similar accounting for their stringent statistical constraints. As for the \texttt{PDT} \textit{ceding} component, the colatitude and azimuthal offset credible intervals only overlap at the 95\% level, with the XMM run inferring the spot to be a bit closer to the south pole. They both have almost the same sizes, with the XMM run preferring slightly larger angular radii. They also share roughly similar temperatures, overlapping at a \CI level, and the posterior of the XMM run moves to higher temperatures. 

Quite a few of these geometric parameters also show correlations with the compactness, particularly the colatitude of the \texttt{CST} region, the size and temperature of its emitting component, the size of its \textit{masking} component, and the colatitude and temperatures of both the \texttt{PDT} components. The size of the \texttt{CST} emitting and \textit{masking} components exhibit especially strong correlations. The various combinations of these parameters and the star's radius could result in the formation of such a bimodal structure in compactness.

For the joint NICER and XMM-Newton run, the inferred NICER background in the lower channels lies almost halfway in between the NICER background limits, whereas the NICER-only run was almost pushing up against the upper limit. For the remaining channels, they appear to be consistent with each other. The inferred background for XMM-Newton MOS1 and MOS2 are highly consistent with the input background data. The residuals for this run also do not exhibit any identifiable structures that would hint at a model deficiency.

\subsection{Effect of radio constraints}
\label{sec: CST+PDT radio free}

The change from PPTA-DR2.5 to the latest PPTA-DR3 radio priors only has relatively minor effects on our overall inferences. We have previously shown that the radius slightly increases upon using the latest priors. We also tested the importance of informing our models with these priors in the context of \jof, by setting up inference runs where we do not use the radio measurements that can only be obtained for binary systems. We attempted these runs both without any background constraints, and with lower and upper limits in place. In both cases, we set the mass to have a uniform prior between 1 and 3 \msol, and a uniform prior over the cosine of the inclination angle between -1 and 1. We still utilize the distance information although this time we do not fix it, but use the actual prior distribution to set up the $\beta$ parameter. We slightly lowered the \ac{X-PSI} resolution (\texttt{max\_sqrt\_num\_cells} = 64, \texttt{num\_energies} = 64) to save on computational expense, especially since accurate parameter estimation is not the primary goal of this run. 

\begin{figure}[t]
    \centering
    \includegraphics[width=0.47\textwidth]{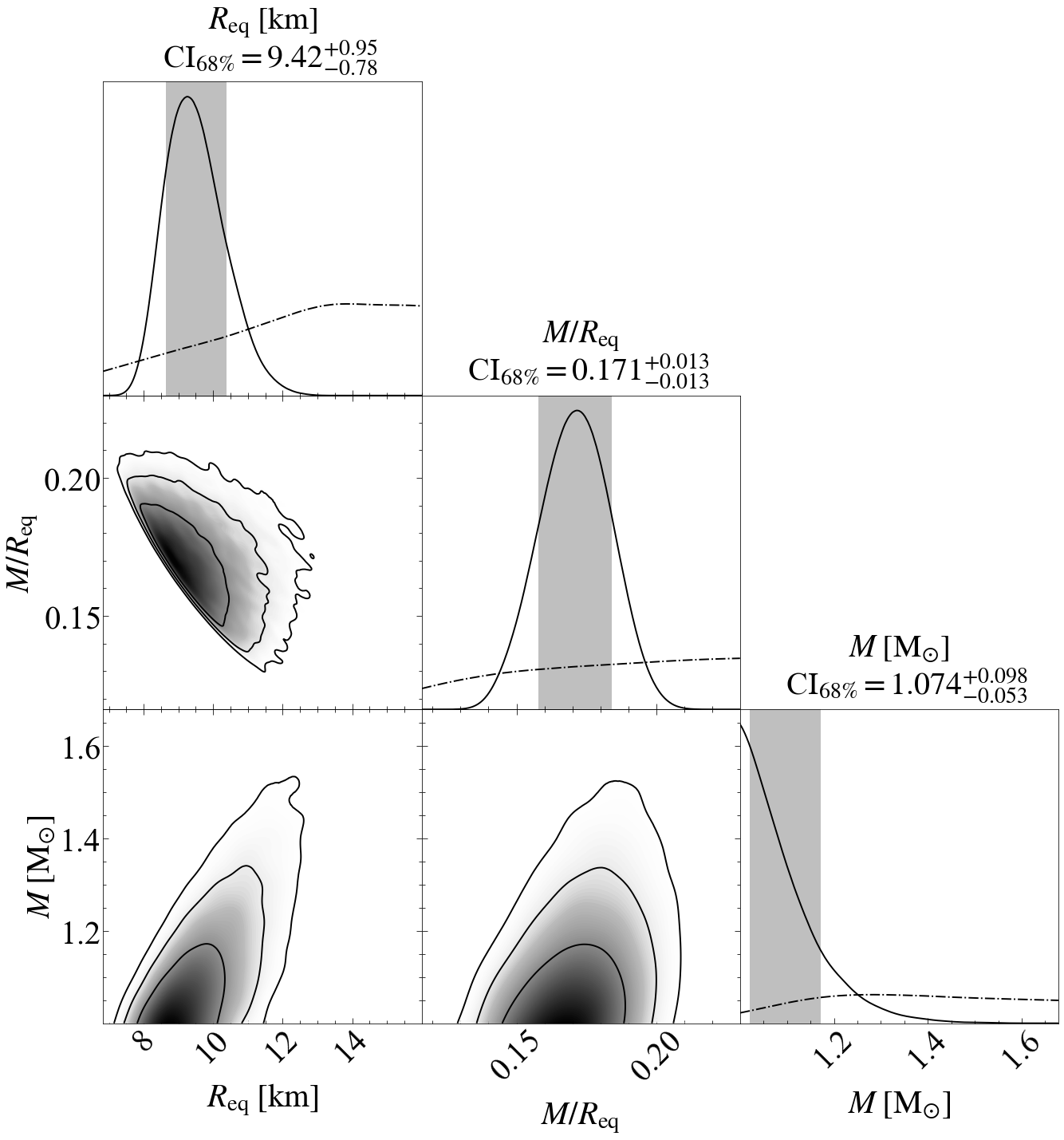}
    \caption{Mass, radius and compactness posteriors inferred by \texttt{CST+PDT} when the models are not informed by radio priors and do not involve any background constraints. In the absence of radio priors, the model prefers much smaller masses.}
    \label{fig: CST+PDT massless}
\end{figure}

With the background constraints in place, the run did not converge in a reasonable time scale. The acceptance fraction reported by \Multinest kept dropping very rapidly as the run progressed. This is not too surprising since the background constraints do not actually reduce the prior hypervolume. Instead, they disfavor many solutions that would previously have had reasonable likelihood values, thus complicating the likelihood surface and making it more challenging for the sampler to correctly identify the relevant modes. We suspect that broadening the prior ranges in light of such highly informative data and in the presence of background constraints likely resulted in an extremely jagged likelihood surface consisting of multiple highly peaked modes, that then turned out to be far too computationally expensive to resolve.

When not using any background constraints, the results are vastly different and do not conform with the radio observations. We obtain a very low mass of $1.074^{+0.098}_{-0.053}$ \msol with the posterior peaking right at the lower edge of our prior bounds (Figure \ref{fig: CST+PDT massless}). Given the extent of the inclination prior space, we infer two separate degenerate modes for inclination caused by reflection symmetry in combination with spot colatitudes. One of these modes is about halfway between the south pole and the equator, much like the radio observations. However, given the tight constraints on both the radio measurement and the inferred posterior mode for inclination, we cannot claim to have accurately recovered the radio observation. The radius that we infer is also much lower than what we have previously obtained using \texttt{CST+PDT} with a \CI of $9.42^{+0.95}_{-0.78}$ km. 

With no constraints in place, the inferred background is once again relatively high with respect to the limits we consider, although with a much larger spread in possible background contributions. The residuals show no discernible structures, indicating that this model can reproduce the data. This run also outperforms the high-res run with no background constraints by 9.228 units in log-evidence. This indicates that while such solutions are far more preferable for explaining the data, our confidence in the radio measurements implies that such physically-viable solutions do not actually represent the true situation. Therefore, for pulse profile modeling of \jof, the radio measurements play a vital role in our inference runs.

\section{Discussion}\label{sec:discussion}

The preferred model that emerges from our analysis, taking into account both upper and lower limits on the background and using NICER data and background estimates, has a mass of $M = 1.418\pm 0.037$ \msol (largely informed by the PPTA-DR3 mass prior) and a radius of $R = 11.36^{+0.95}_{-0.63}$ km (68\% credible intervals).  The associated hot spot geometry has a large single temperature ring encompassing the north pole, and a dual temperature spot (consisting of a small high temperature component overlapping a larger low temperature component) between the equator and the south pole. The observer views the star from the southern hemisphere at an inclination angle of $137.506 \pm 0.019$ degrees. This inclination angle does not match the $16^\circ$ inferred from modeling the position angle of the linearly polarized part of PSR~J0437$-$4715's radio emission \citep{Gil97}, however see Section~\ref{subsec:surf_heat} for a discussion of the limitations of the radio data analysis.

\subsection{Modeling assumptions}

Our modeling is conditional upon the tight mass prior obtained from radio timing. In the analysis of \jos by \citet{Riley21}, the impact of the mass prior provided by radio timing was assessed; without it, no constraining mass posterior was recovered (the posterior changed very little from the unconstrained mass prior) -- it would not have been concluded independently that it was a high mass pulsar. For \jof, when we drop the informative mass prior, we find that the recovered posterior does differ substantially from the broad prior but it disfavors the known mass very strongly (at least when not imposing any background constraints). Unfortunately due to computational cost we have not yet been able to determine whether the imposition of background constraints would change this. Interestingly, for \joo, which does not have a mass prior since it is not in a binary, simulations have demonstrated that it is nevertheless possible to recover the injected mass \citep{Vinciguerra23a}. The factors determining reliance on an informative mass prior are clearly something that merits further detailed investigation, since several of the NICER PPM targets lack mass priors. 

In our modeling we have made use of available background constraints: when analyzing NICER data alone, we used both 3C50 instrument background estimates and an estimate of the AGN contribution to put lower and upper limits on the background. This does make a difference to our results:  without background constraints we would have inferred a slightly lower radius of $11.01^{+0.51}_{-0.48}$ km.  Including XMM-Newton data in joint modeling provides a useful additional check: there the solution is somewhat bimodal, but results are consistent with the NICER-only analysis. The fact that background constraints have an effect shows the importance of developing good instrumental background models and also monitoring variable sources in the field of view.

The AGN is variable on a range of timescales. \citet{HalpenMarshall96} observed it for 20 days with EUVE in the 0.1--0.2 keV band and found that it varied unpredictably with as short as a 5 hour doubling time and a total range of a factor of 4 during the full span. We also looked at a 110 ks XMM-Newton observation from 2009 December 15 (ObsID 0603460101) 
and again saw rapid variability of about 50\% on 10 ks timescales. This means that only fully simultaneous observations would be sufficient to accurately measure the AGN contribution to the observed NICER spectrum, but this is unfeasible for the large number of short observations that are a consequence of NICER's visibility constraints on the ISS.

Our primary result uses a fully ionized hydrogen atmosphere.
As shown in \citet{Salmi23}, the atmosphere choices, such as hydrogen versus helium composition, have the potential to significantly affect the inferred neutron star parameters, including radius.
However, not all choices are equally probable or able to explain the data.
Hydrogen composition is considered most likely due to the rapid sinking of heavier elements \citep{Alcock80,Hameury83,Brown02,ZP2002}. The tight mass and inclination priors for \jof may also help to reduce the sensitivity to the atmosphere choices, similar to the case of \jos \citep{Salmi23}.

Finally, the present analysis neglected the possible contribution of the non-thermal emission to the phase-resolved signal from \jof. It has been known for some time that this pulsar exhibits non-thermal emission, which had been detected in Chandra data as a power-law tail \citep{zavlin2002}, then confirmed in the XMM-Newton phase-averaged spectrum \citep{bogdanov2006,Bogdanov13}. One possible explanation for the origin of this non-thermal emission is inverse Compton scattering of the thermal X-ray photons by relativistic electrons, located above the polar caps \citep{bogdanov2006,Bogdanov13}. When fitted with a power-law, this non-thermal emission dominates over the thermal emission above $\sim$ 3 keV. However, the XMM-Newton data only could not precisely constrain the slope of the power-law because it was largely correlated with the choices of model for the emission at lower energies (below 3 keV). Observations with NuSTAR in the 3--20~keV energy range provided a more robust estimate of power-law ($\Gamma\sim 1.60\pm0.25$, 90\% confidence interval) when using the NuSTAR alone, i.e., free of correlation with the low-energy emission \citep{Guillot16}.

The NuSTAR observation also revealed that the hard X-ray, non-thermal, emission was also pulsed at the frequency of the pulsar (a detection at $3.7\sigma$ \footnote{This detection significance was revised to $\sim 7\sigma$ (Guillot et al., in prep.) after clock corrections were provided by the NuSTAR team \citep{bachetti21}. } \citealt{Guillot16}). While this pulsed hard X-ray emission is present above 3~keV, i.e., outside of the energy range of the NICER data analysed in the present letter, its contribution could extend to lower energies and account for a small fraction (of the order of a few percent) of the phase-resolved signal from the pulsar. However, since the mechanism responsible for the non-thermal emission remains unknown, it is difficult to assert the energy range where one might expect the pulsed non-thermal emission to be present. For example, inverse Compton scattering of thermal X-ray photons by a hot plasma of electrons above the polar caps could result in pulsed hard X-ray emission, but it would mostly be above $\sim 3$~keV \citep[see Figure 2 in ][for Comptonized blackbody emission]{bogdanov2006}. In the absence of a firm identification of the origin and energy extent of the pulsed non-thermal emission, we have for now neglected any potential contribution to the phase-resolved signal.

\subsection{Radius measurement in context}

In this section we compare our results for the radius of a $\sim 1.4$ \msol neutron star with those obtained from other studies or techniques that do not involve the specification of an EoS model. 

There are several previous studies that have attempted to place constraints on the radius of PSR J0437--4715.  Some early work attempting to fit radiative profiles and constrain polar cap properties and the mass to radius ratio (before a precision mass measurement was available) was carried out by \citet{Pavlov97,Zavlin98}. Once a mass measurement was available, \citet{Bogdanov13} modeled the pulse profile of the XMM-EPIC dataset, assuming two single temperature circular hot spots and concluding that $R>11.1$ km (at the $3\sigma$ confidence level) for the then current radio timing mass measurement of 1.76 \msol. Using the current mass measurement would have reduced the lower limit on the radius.  Our lower limit ($R > 10.5$ km at the 68\% level) is smaller than the values from \citet{Bogdanov13}, however given the improvement in data quality and modeling, we do not find this difference surprising. 

More recently \citet{Gonzalez19} carried out joint modeling of phase-averaged far ultraviolet (FUV) and soft X-ray data for \jof, using data from the Hubble Space Telescope (HST) and R\"{o}ntgensatellit (ROSAT, in the 0.1--0.3\,keV range), respectively.  They inferred a radius of $13.6^{+0.9}_{-0.8}$\,km, higher than the value inferred in this current study.  That analysis modeled the $\sim10^{5}$\,K emission from the rest of the surface (outside the hot spots) using atmosphere models compatible with that range of temperatures, for hydrogen, helium and iron compositions. Using fixed values for the pulsar mass and distance and priors on the amount of extinction $E\left(B-V\right)$ for the FUV data, the authors favored a hydrogen composition of the atmosphere, although a helium composition was not fully excluded. In the final results of \citet{Gonzalez19}, the radius was obtained with a model that included the phase-average contributions of the hot spots to the spectra, assuming a blackbody emission for these and using parameter values from the work of \citet{Guillot16}.  Although the spots have minor contributions to the total phase-averaged spectrum below 0.3\,keV, this approach ensured that this contribution would not bias the spectral fits of the rest of the surface at $\sim10^{5}$\,K. 

However, an update to the HST calibration revealed that the absolute flux of the HST Advanced Camera for Surveys (ACS) Solar Blind Channel (SBC) was incorrect. Specifically, the zero-point flux is actually lower than previously thought by $\sim30\%$, thus overestimating the measured FUV fluxes\footnote{See \url{https://archive.stsci.edu/contents/newsletters/october-2019/sbc-absolute-flux-calibration-revised-by-30}}.  These data, comprising the bulk of the FUV data points in \citet{Gonzalez19} likely biased the radius measurement, possibly explaining the difference in radius with the one in the present work. In an effort to mitigate this bias, a re-analysis of the FUV + X-ray spectrum of \jof was performed (Stammler et al., in prep) excluding the ACS-SBC data, and only using data from the HST Space Telescope Imaging Spectrograph FUV Multi-Anode MicroChannel Array for which the calibration has not been revised. In addition, this new analysis used an updated, much tighter prior than previously on the extinction $E\left(B-V\right)$, obtained from the GAIA datasets \citep{Vergely2022}. The full priors\footnote{In that analysis, the priors used were: $E(B-V)=0.005\pm0.003$, $M=1.44\pm0.07$~\msol and $d=157\pm0.2$~pc} on the mass and on the source distance were also used in that recent analysis, instead of fixed values. Finally, the phase-averaged contribution of the hot spots was modeled realistically with atmosphere models, instead of assuming blackbody emission. Overall, the new, updated, radius of \jof measured from the FUV and soft X-ray data, is $12.3\pm0.9$\,km (Stammler et al., in prep.), which is compatible with the radius presented in this work.

We can also compare the mass and radius inferred for \jof to the masses and radii inferred for other NICER sources.
For the heavy pulsar \jos, the current best X-PSI result\footnote{See \citet{Dittmann24} for analysis using an independent PPM pipeline for this source.} is $M = 2.073_{-0.069}^{+0.069}$ \msol and $R = 12.49_{-0.88}^{+1.28}$ km \citep{Salmi24}.  This means that $\Delta R = R_{2.0} - R_{1.4}= 1.13_{-1.08}^{+1.59}$ km.  The implications of this change in radius over this mass range are discussed in more detail in Section \ref{subsec:eos}.

For \joo the picture is more complex \citep{Vinciguerra23b}:  the current best NICER-only (with no background constraints) results are $M = 1.37 \pm 0.17$ \msol, $R = 13.11 \pm 1.30$ km for the \texttt{ST+PST} model, consistent with the earlier results of \citet{Riley19}\footnote{See also \citet{Miller19} for analysis using an independent PPM pipeline for this source.}.  However when taking into account XMM-Newton data to provide constraints on background, two different solutions are preferred, with masses and radii of $[M = 1.40^{+0.13}_{-0.12} ~\mathrm{M_\odot}, R = 11.71^{+0.88}_{-0.83}$ km] and $[M = 1.70^{+0.18}_{-0.19} ~\mathrm{M_\odot}, R = 14.44^{+0.88}_{-1.05}$ km], depending on the assumed surface pattern model \citep{Vinciguerra23b}.  The mass and radius for the first mode identified in the joint NICER-XMM analysis are extremely consistent with our findings for \jof; the NICER-only solution (which also has a similar mass but a higher radius) less so.  The higher mass, higher radius NICER-XMM mode of \joo (although favored according to the evidence calculation) is also harder to reconcile with our new findings. 

Note that while rotation does increase the equatorial radius of a neutron star (and the resulting oblateness is incorporated into our ray-tracing models), the change for a 1.4 \msol neutron star as one goes from zero rotation to 174 Hz is at most $\sim 0.2$ km (for the stiffest EoS with the largest radii), see Figure 5 of \citet{Raaijmakers19}.  For lower radii, such as we infer for \jof, the difference is much smaller than our inferred credible intervals.  

\subsection{Equation of State implications}
\label{subsec:eos}

A full EoS analysis of the implications of our mass and radius for \jof is presented in a companion letter by \citet{Rutherford24}.  In this Section we therefore highlight only a few key points. 

\begin{figure}[t]
    \centering
    \includegraphics[width=0.47\textwidth]{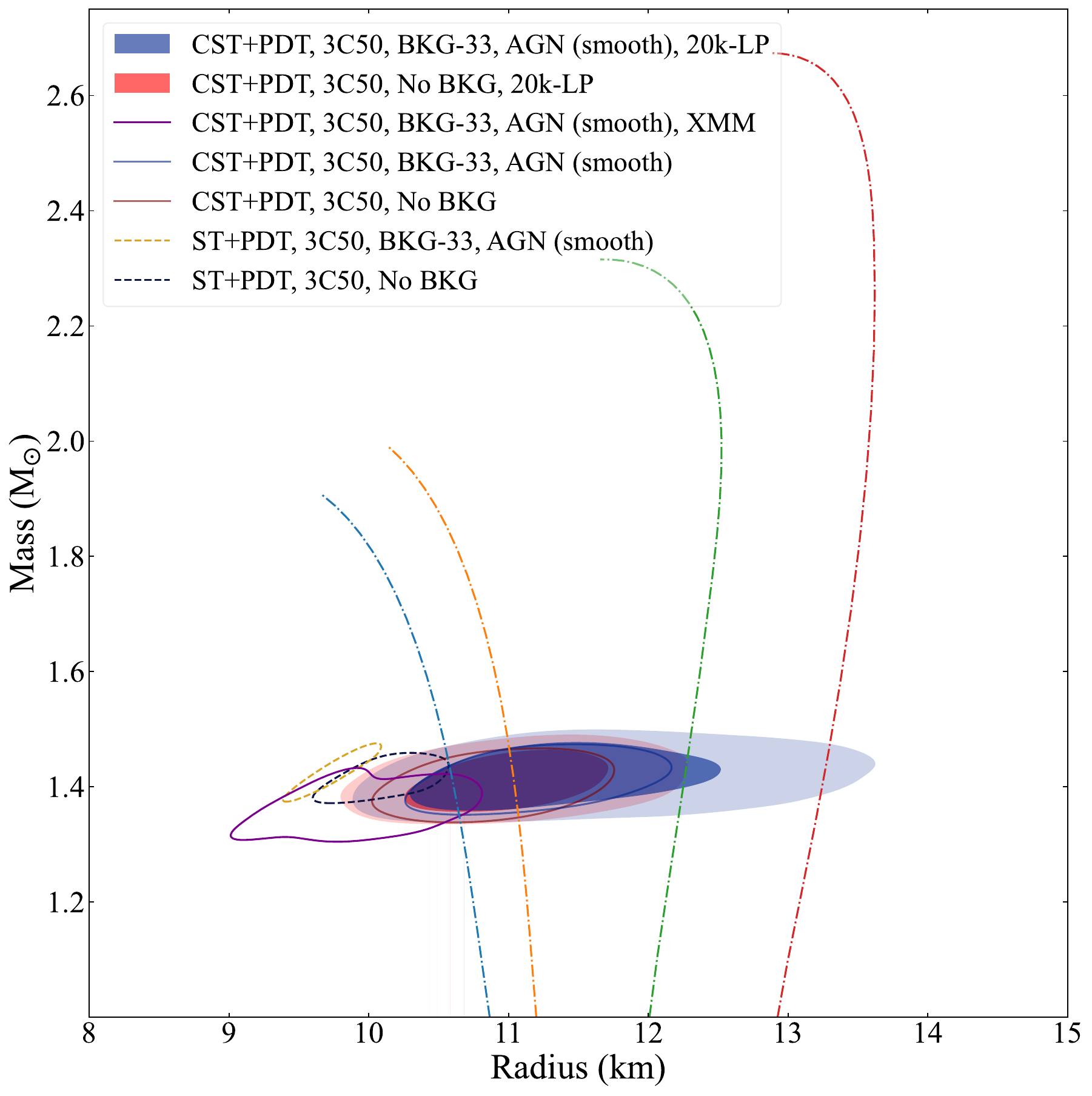}
    \caption{Inferred mass and radius posteriors for the different capable models explored in this paper, compared to mass-radius curves from EoS models. The shaded contours are the 68\% and 95\% credible regions for the high-res \texttt{CST+PDT} runs, and the remaining contours represent the 68\% credible regions for the default resolution \texttt{CST+PDT} and \texttt{ST+PDT} runs. The colored dash-dot lines are example mass-radius curves generated using one of the families of EoS models explored in \citet{Rutherford24}. It employs the Baym-Pethick-Sutherland crust EoS \citep{Baym71} at densities less than 0.5 times the nuclear saturation density $n_s$; the N$^3$LO $\chi$EFT band from \citet{Keller2023} at densities 0.5-1.5 $n_s$; and a parametrized piecewise polytropic (PP) model \citep{Hebeler2013} at higher densities.  The parameters for the mass-radius relations shown here have been chosen to span the range of stiffness allowed by this model that is also compatible with the existence of $\sim 2.0$ \msol neutron stars.}
    \label{fig: MR EoS}
\end{figure}

The radius that we infer for \jof favors softer EoS models in the vicinity of the nuclear saturation density $n_s$. Figure \ref{fig: MR EoS} shows the comparison to several example mass-radius curves from the EoS models explored in \citet{Rutherford24}.

The radius inferred for \jof is also more consistent with the results that have emerged to date from gravitational wave measurements of tidal deformabilities. Using EoS insensitive relations, \citet{GW170817_TD1} reported masses and radii for the two neutron stars involved in the binary merger GW170817 of [$M_1 = 1.15-1.36$ \msol, $R_1 = 10.8^{+2.0}_{-1.7}$ km] and [$M_2 = 1.36 - 1.62$ \msol, $R_2 = 10.7^{+2.1}_{-1.5}$ km] (90\% credible intervals). When using an EoS model their inferred masses and radii were [$M_1 = 1.18-1.36$ \msol, $R_1 = 11.9^{+1.4}_{-1.4}$ km] and [$M_2 = 1.36 - 1.58$ \msol, $R_2 = 11.9^{+1.4}_{-1.4}$ km].  Meanwhile \citet{Raaijmakers2021}, using different EoS models, inferred $R_{1.4} = 11.51^{+1.51}_{-1.47}$ km (PP model, 95\% credible intervals) and $R_{1.4} = 11.18^{+1.33}_{-1.51}$ km  (CS model) when using only the tidal deformabilities from GW170817 and GW190425.  

Our results appear less consistent with the neutron star radii that have been inferred from the results of the parity-violating electron scattering experiment PREX-2 \citep{PREX}. PREX measures the neutron skin thickness of $^{208}$Pb, which constrains the density dependence of the symmetry energy, with the results of \citet{PREX} implying a very stiff symmetry energy. Given the expected correlation between symmetry energy and neutron star radius, \citet{Reed21} have argued that this would imply $R_{1.4} \gtrsim 13.25$ km. However the PREX-2 uncertainties are broad \citep{Reed21} and there are also apparent tensions with more recent measurements of the neutron skin thickness of $^{48}$Ca taken by the CREX experiment \citep{CREX} and the Large Hadron Collider (LHC) inference of the neutron skin thickness of $^{208}$Pb \citep{Giacalone23}.
In fact, if one treats the results from the PREX and CREX experiments with equal weight, one finds that only nuclear interactions with symmetry energy parameters $J=32.2\pm1.7$ MeV and $L=52.7\pm13.2$ MeV \citep{lattimer-particles} can accommodate both experiments simultaneously ($J$ and $L$ being the symmetry energy and its slope at nuclear saturation density). In addition, the LHC measurement implies $L\simeq70\pm40$ MeV \citep{Sun23}.
In comparison, nuclear mass fitting with a wide variety of both non-relativistic and relativistic nuclear interactions yields $J=32.5\pm2.1$ MeV and $L=57.7\pm20.7$ MeV.  Moreover, combining Chiral Effective Field Theory ($\chi$EFT) neutron matter calculations with estimates of the nuclear saturation window ($n_s=0.155\pm0.010$ fm$^{-3}$ and symmetric matter binding energy at saturation $B=16.1\pm0.3$ MeV) gives $J=32.0\pm1.1$ MeV and $L=51.9\pm7.9$ MeV. These estimates are in remarkable agreement with one another. Neutron star models constructed with those interactions that best fit both PREX and CREX data imply $R_{1.4}=11.6\pm1.0$ km \citep{lattimer-particles}. Thus, nuclear experiment and theory are very consistent with our results for the radius of \jof.

The radius measurement of \jof, combined with that for \jos \citep{Salmi24}, implies that the radius difference between 2.0\msol and 1.4\msol stars \citep[a quantity whose significance was first discussed by][]{Drischler21}, $\Delta R = R_{2.0} - R_{1.4} \simeq 1.13^{+1.59}_{-1.08}$ km.  
The central value for $\Delta R$ is extremely large (although the uncertainties are also large). For comparison, \citet{Sun23}'s compilation of 313 non-relativistic and relativistic equations of state that satisfy $M_{\rm max} \ge 2.0$\msol, where $M_{\rm max}$ is the maximum possible neutron star mass, have one EoS model with $\Delta R \ge 1.0$ km, and only 4\% have $\Delta R \ge 0.5$ km.  \citet{Drischler21} states that if $\Delta R > 0$ km, standard extrapolations would predict some unusual stiffening (including, perhaps, a quarkyonic or crossover transition to quark matter) should occur below $2.6 - 2.8 n_s$.  
Furthermore, if $\Delta R$ turns out to be greater than 0.5 km, then one should expect that this stiffening would not only occur for densities $\le 2n_s$, but also would be accompanied by a very high value for $M_{\rm max}$, one that could be in conflict with expectations from the neutron star merger GW170817 and in particular modeling of its electromagnetic counterpart ($M_{\rm max} \lesssim 2.3$\msol, \citealt{Margalit2017, Rezzolla18, Shibata19, Lattimer21}).

\subsection{Surface heating configuration}
\label{subsec:surf_heat}

The favored hot spot configuration identified by our analysis, shown in Figure \ref{fig: CST+PDT projection}, is not consistent with a centered dipole magnetic field but could potentially be explained by an offset dipole magnetic field or a field with quadrupole components. Via magnetospheric modeling, \citet{Lockhart19} (building on the work by \citealt{Gralla17}) infer similar surface emission patterns for \jof, consisting of a quadrudipolar magnetic field. While there are some dissimilarities to our inferred geometry, these likely stem from the more restrictive fitting procedure, with more fixed model neutron star parameters, used in their analysis.

These findings are also consistent with the picture from radio observations: \jof has complex radio pulse morphology that is challenging to model assuming a dipolar magnetic field and plausible radio emission microphysics \citep{Bhat14}. The swing of position angle (PA) of linearly polarized radio emission does not conform to the prescriptions of a widely used phenomenological rotating vector model, which again relies on the dipole field assumption \citep{Manchester95,Navarro97}. Some complexities of the PA behavior may arise from mixing between two orthogonally polarized modes. However a detailed study of \jof’s single-pulse emission, which resolved the modes, showed that mode mixing can not explain all peculiarities of PA swing \citep{Oslowski14}. It has been supposed therefore that radio emission from \jof originates either close to surface where higher-order multipoles are present \citep{Oslowski14}, or closer to the light cylinder where large-scale magnetic field is modified by magnetospheric currents \citep{Gil97}. 

The fact that the field configuration appears to be more complex than a simple centered dipole is consistent with findings for the other two NICER MSPs analyzed to date.  Initial analysis of \joo NICER data indicated a highly non-dipolar field structure and complex emitting region shapes \citep{Riley19,Bilous_2019,Miller19}. More recent joint analysis of NICER and XMM data suggests that a more antipodal geometry could be possible, but with the hot regions having a more complex dual temperature distribution \citep{Vinciguerra23b}.  The heavy pulsar \jos also has a definitively non-antipodal hot region geometry: \citet{Salmi22, Salmi24} found that the offset angle from the antipode is above $25^\circ$ with 84\% probability. 

Neutron stars are initially formed with complex magnetic field structures as a result of supernova explosions \citep{Ardeljan2005, Obergaulinger2017}. These fields are further modified  over time by various diffusive processes \citep{Reisenegger2009, Vigano2013, Mitchell2015, Gourgouliatos2018, Igoshev2023}. It remains uncertain whether a complex multipolar field can persist for long periods in isolated neutron stars, but the field can also be altered during the accretion-induced spin-up phase that produces rotation-powered millisecond pulsars \citep{Romani1990, Melatos2001, Payne2004}. Although quantitative modeling of hot spots is still pending, non-antipodal hot spots are expected as a natural outcome of accretion-induced magnetic burial \citep{Suvorov2020}. Accurate modeling of hot spot shapes and locations must also incorporate the coupled magneto-thermal evolution of the crust: in isolated neutron stars, hot spots can be displaced from their antipodal positions by past heat deposition events, even in a dipole-dominated magnetic field, provided that the crustal transport properties are properly considered \citep{deGrandis2020}.

\section{Conclusions}\label{sec:conclusions}

In this letter, we have presented NICER's first mass, radius and geometry inferences for the bright $\sim 1.4$ \msol pulsar \jof, using pulse profile modeling of NICER XTI X-ray spectral-timing data, with priors on mass, distance and inclination from PPTA radio pulsar timing data.  For our preferred model, we find an inferred mass $M = 1.418 \pm 0.037$ \msol (dominated by the radio prior) and radius of $R = 11.36^{+0.95}_{-0.63}$ km.  This radius favors softer dense matter equations of state and is highly consistent with constraints derived from gravitational wave measurements of neutron star binary mergers.  

The information from the radio prior is critical - our (limited) tests indicate that without it we would not have recovered the correct mass. Our results are also dependent on having good estimates for the astrophysical background, including a contribution from a bright AGN in the NICER field of view. Better monitoring of the AGN may enable tighter background constraints in the future, and NICER continues to build up a larger dataset on \jof that will also enable tighter constraints on radius.

PPM also allows us to constrain the geometry of the X-ray emitting hotspots on the stellar surface, which are assumed to be the heated magnetic poles.  The configuration that we find is not consistent with a centered dipole magnetic field, but could be consistent with either an offset dipole or the presence of quadrupole components. This is consistent with the picture from radio and $\gamma$-ray modeling of this pulsar, which has long pointed to a complex magnetic field structure.

\begin{acknowledgments}
This work was supported in part by NASA through the NICER mission and the Astrophysics Explorers Program. 
D.C., T.S., S.V., T.E.R., Y.K., A.L.W. and B.D. acknowledge support from ERC Consolidator Grant No.~865768 AEONS (PI: Watts). 
S.B.~acknowledges funding from NASA grants 80NSSC20K0275 and 80NSSC22K0728. 
N.R. was supported by NASA Grant No.~80NSSC22K0092.
S.G. acknowledges the support of the CNES.
S.M. acknowledges support from NSERC Discovery Grant RGPIN-2019-06077.
W.C.G.H. acknowledges support through grant 80NSSC23K0078 from NASA.
Parts of this research were conducted by the Australian Research Council Centre of Excellence for Gravitational Wave Discovery (OzGrav), through project numbers CE170100004 and CE230100016.
The use of the national computer facilities in this research was subsidized by NWO Domain Science.  
Part of the work was carried out on the HELIOS cluster including dedicated nodes funded via the aforementioned ERC CoG.
Astrophysics research at the Naval Research Laboratory is supported by the NASA Astrophysics Explorer Program.
We acknowledge extensive use of NASA’s Astrophysics Data System (ADS) Bibliographic Services and the ArXiv.
\end{acknowledgments}

\software{\texttt{X-PSI} \citep{Riley23}, GNU Scientific Library (\texttt{GSL}; \citealt{Gough2009}), \texttt{HEASoft} \citep{heasoft2014}, \texttt{MPI} for Python \citep{Dalcin2008}, \Multinest \citep{Feroz09}, \Pymultinest \citep{PyMultiNest}, \texttt{nestcheck} \citep{Higson2018JOSS}, \texttt{GetDist} \citep{Lewis2019}, \texttt{Jupyter} \citep{2007CSE.....9c..21P, kluyver2016jupyter}, \texttt{astropy
} \citep{astropy:2013, astropy:2018, astropy:2022}, \texttt{scipy} \citep{2020SciPy-NMeth, scipy_10909890}, \texttt{matplotlib} \citep{Hunter:2007}, \texttt{numpy} \citep{numpy}, \texttt{python} \citep{python} and \texttt{Cython} \citep{cython:2011}.}
Software citation information aggregated using \texttt{\href{https://www.tomwagg.com/software-citation-station/}{The Software Citation Station}} \citep{software-citation-station-paper, software-citation-station-zenodo}.

\bibliographystyle{aasjournal}
\bibliography{allbib}

\clearpage 
\appendix
\section{CST+PDT results table}
\begin{longtable}{l|l|l|ll}

\caption{Summary table for the headline high-res CST+PDT run described in Section \ref{sec: Prod analyses}}\label{tab: CST+PDT}\\
\hline\hline

Parameter & Description & Prior (density and support) & $\widehat{\textrm{CI}}_{68\%}$ & $\widehat{\textrm{ML}}$\\
\hline
\endfirsthead
\multicolumn{5}{c}%
{\tablename\ \thetable\---\textit{Continued from previous page}} \\
\hline
\endhead
\hline \multicolumn{5}{r}{\textit{Continued on next page}} \\
\endfoot
\hline\hline
\endlastfoot
$P$ $[$ms$]$ &
coordinate spin period &
$5.7575$, fixed & 
$-$ &
$-$ \\
\hline
$M$ $[\textrm{M}_{\odot}]$ &
gravitational mass &
$N(1.418,0.044^{2}) $ &
$1.418_{-0.037}^{+0.037}$ &
$1.388$ \\
\hline
$R_{\textrm{eq}}$ $[$km$]$ &
coordinate equatorial radius &
$U(3r_{\rm g}(1),16)$ &
$11.36_{-0.63}^{+0.95}$ &
$13.17$ \\
\hline
&compactness condition & $13.7\leq \log_{10}g_{\textrm{eff}}(\theta)\leq15.0$,~$\forall\theta$\\
\hline
$\Theta_{p}$ $[$radians$]$&
$p$ region center colatitude &
$\cos(\Theta_{p})\sim U(-1,1)$ &
$0.146_{-0.022}^{+0.028}$ &
$0.112$ \\
$\Theta_{s}$ $[$radians$]$&
$s$ superseding component center colatitude &
$\cos(\Theta_{s})\sim U(-1,1)$ &
$2.348_{-0.070}^{+0.060}$ &
$2.392$ \\
$\Theta_{c,s}$ $[$radians$]$&
$s$ ceding component center colatitude &
$\cos(\Theta_{c,s})\sim U(-1,1)$ &
$2.307_{-0.095}^{+0.073}$ &
$2.318$ \\
$\phi_{p}$ $[$cycles$]$ &
$p$ region initial phase\footnote{With respect to the meridian on which Earth lies.} &
$U(-0.25,0.75)$, wrapped &
$0.4429_{-0.0062}^{+0.0054}$ &
$0.4381$ \\
$\phi_{s}$ $[$cycles$]$ &
$s$ superseding component initial phase\footnote{With respect to the meridian on which the Earth antipode lies.} &
$U(-0.25,0.75)$, wrapped &
$0.4704_{-0.0036}^{+0.0040}$ &
$0.466$ \\
$\chi_{s}$ $[$radians$]$ &
azimuthal offset between the $s$ components&
$U(-\pi,\pi)$ &
$0.0041_{-0.0133}^{+0.0160}$ &
$0.0121$ \\
$\zeta_{p}$ $[$radians$]$ &
$p$ emitting component angular radius &
$U(0,\pi/2)$ &
$0.433_{-0.060}^{+0.061}$ &
$0.561$ \\
$\zeta_{o,p}$ $[$radians$]$ &
$p$ omitting component angular radius &
$f \sim U(0,1)$, $f: =\zeta_{o,p}/\zeta_{p}$&
$0.139_{-0.076}^{+0.102}$ &
$0.336$ \\
$\zeta_{s}$ $[$radians$]$ &
$s$ superseding component angular radius &
$U(0,\pi/2)\footnote{\label{footnote: cede-supercede}$\zeta_{s}$, $\zeta_{c,s}$ are functions of $\Theta_{s}$, $\Theta_{c,s}$, $\phi_{s}$, $\chi_{s}$ such that superseding component never engulfs ceding component.}$ &
$0.0299_{-0.023}^{+0.003}$ &
$0.026$ \\
$\zeta_{c,s}$ $[$radians$]$ &
$s$ ceding component angular radius &
$U(0,\pi/2)\footref{footnote: cede-supercede}$ &
$0.197_{-0.017}^{+0.020}$ &
$0.177$ \\
\hline
&non-overlapping hot regions & function of \\
& & $(\Theta_{p}, \Theta_{s}, \Theta_{c,s}, \phi_{p}, \phi_{s}, \chi,$ \\
& & $\zeta_{p}, \zeta_{o,p}, \zeta_{s}, \zeta_{c,s})$\\
\hline
$\log_{10}\left(T_{p}\;[\textrm{K}]\right)$ &
$p$ region effective temperature &
$U(5.1,6.8)$, limits &
$6.101_{-0.015}^{+0.013}$ &
$6.075$ \\
$\log_{10}\left(T_{s}\;[\textrm{K}]\right)$ &
$s$ superseding component effective temperature &
$U(5.1,6.8)$, \TT{NSX} limits &
$6.219_{-0.013}^{+0.013}$ &
$6.199$ \\
$\log_{10}\left(T_{c,s}\;[\textrm{K}]\right)$ &
$s$ ceding component \TT{NSX} effective temperature &
$U(5.1,6.8)$, \TT{NSX} limits &
$5.752_{-0.028}^{+0.026}$ &
$5.724$ \\
$\cos(i)$ &
cosine Earth inclination to spin axis &
$\cos(N(2.39993,0.00028^{2})) $ &
$-0.7373_{-0.0002}^{+0.0002}$ &
$-0.7376$ \\
$D$ $[$kpc$]$ &
Earth distance &
$0.15698$, fixed & 
$-$ &
$-$ \\
$N_{\textrm{H}}$ $[10^{20}$cm$^{-2}]$ &
interstellar neutral H column density &
$U(0.004,2.0)$ &
$0.05_{-0.04}^{+0.68}$ &
$0.02$ \\
$\alpha_{\rm{XTI}}$ &
NICER effective-area scaling &
$N(1,0.1^{2})$ &
$1.013_{-0.079}^{+0.079}$ &
$0.979$ \\
\hline\hline
&Sampling process information&&& \\
\hline
&number of free parameters:\footnote{In the sampling space; the number of background count rate variables is equal to the number of channels defined by the NICER data set.} $18$ &&& \\
&number of processes:\footnote{The mode-separation MultiNest variant was deactivated, meaning that isolated modes are not evolved independently and nested sampling threads contact multiple modes.} $1$ &&& \\
&number of live points: $2\times10^{4}$ &&& \\
&sampling efficiency: $0.3$ &&& \\
&termination condition: $10^{-1}$ &&& \\

\end{longtable}

\newpage
\section{Supplementary corner plots of posteriors inferred for different model parameters by exploratory runs} \label{appendix: figs}

\begin{figure*}[h]
    \centering
    \includegraphics[width=\textwidth]{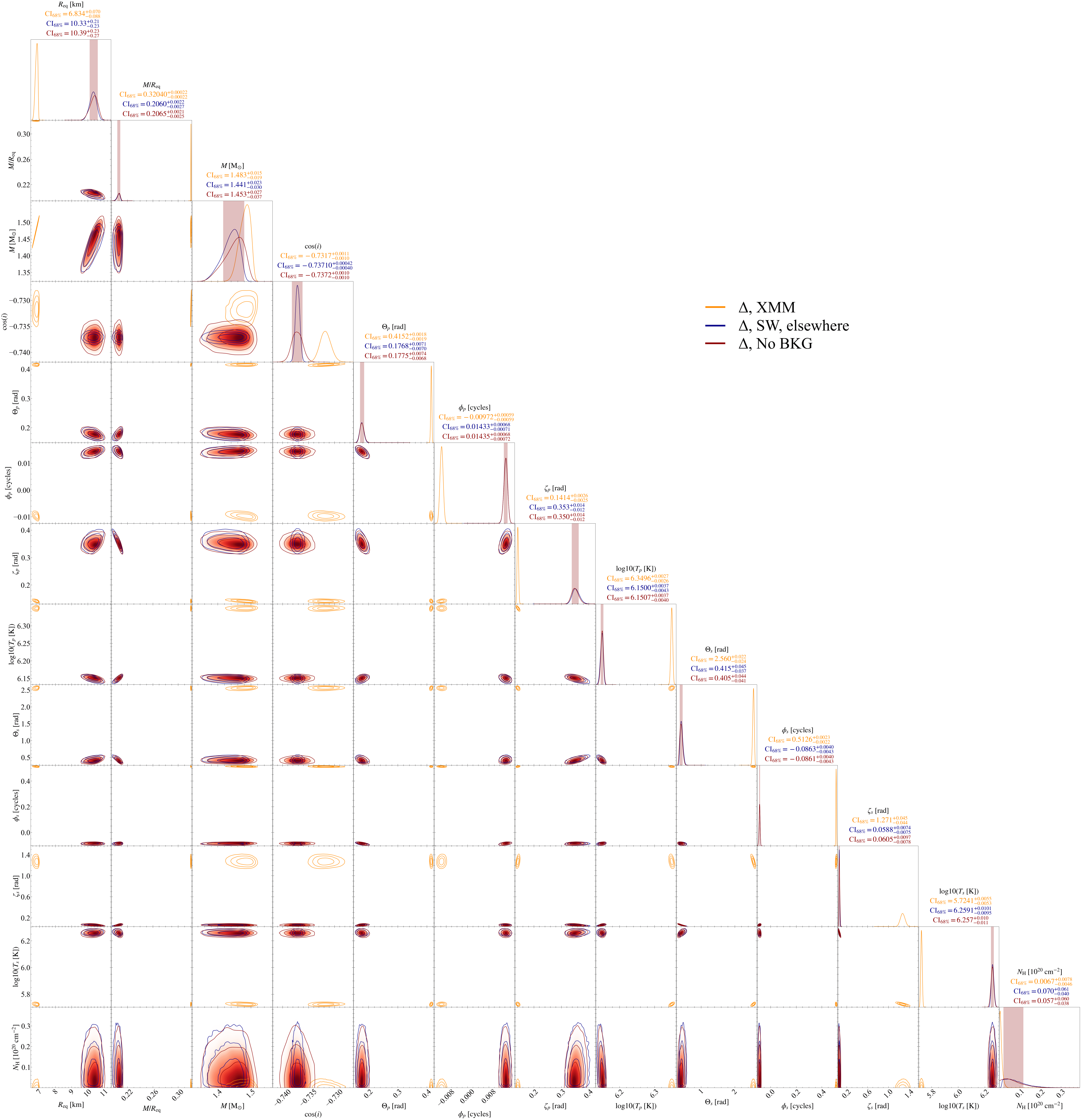}
    \caption{1-D and 2-D marginalized posteriors inferred by \texttt{ST-U} when applied to the Delta ($\Delta$) dataset, using either no background, Space Weather (SW) + \texttt{elsewhere} component, or informed by XMM-Newton (data and background).}
    \label{appendixfig: ST-U delta all params}
\end{figure*}

\begin{figure*}[h]
    \centering
    \includegraphics[width=\textwidth]{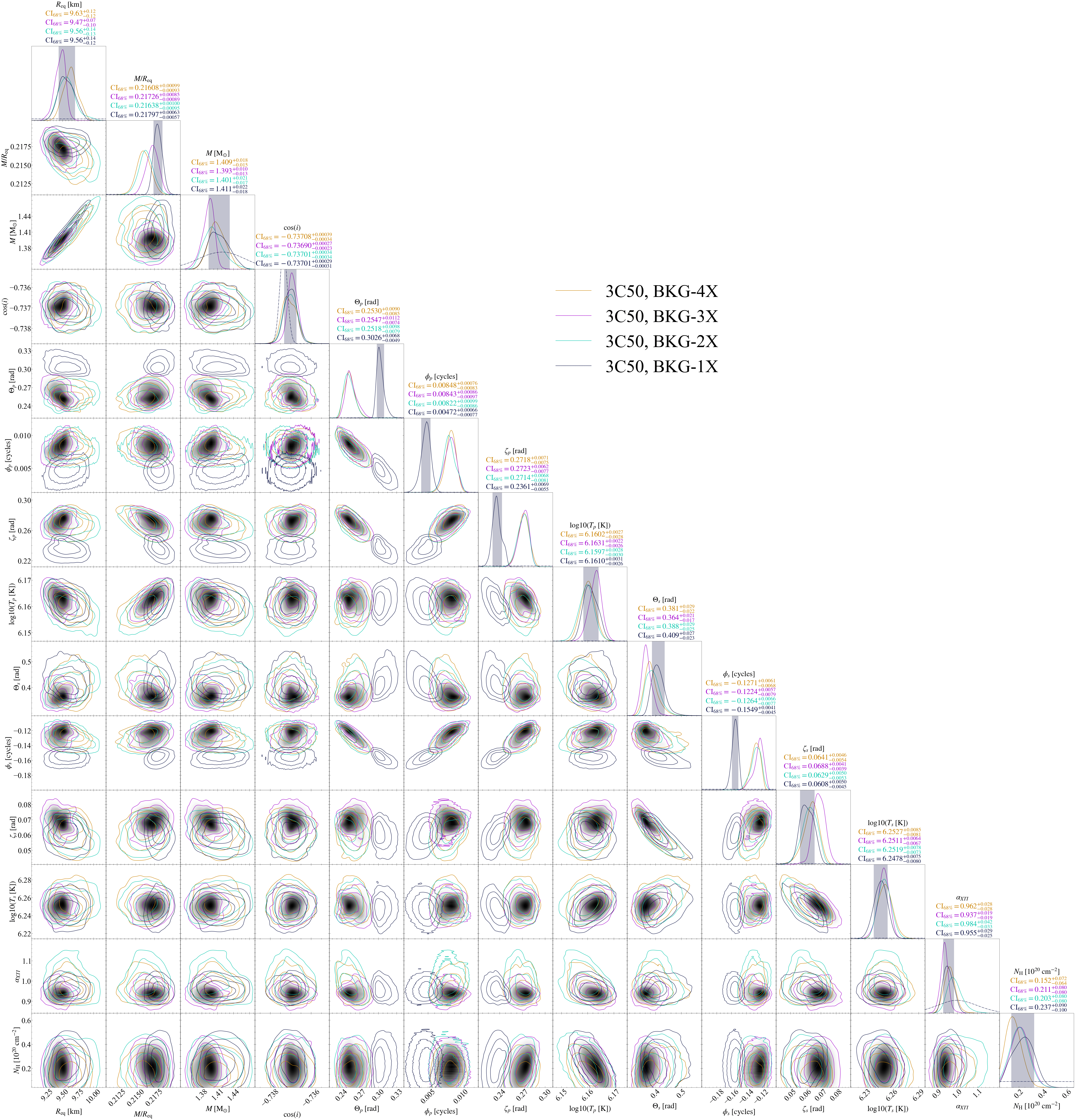}
    \caption{Posteriors obtained by \texttt{ST-U} as applied to the 3C50 dataset, testing the implementation of different $n_{l}$ values when applying a lower 3C50 instrument-only background constraint.}
    \label{appendixfig: ST-U 3C50 uncertainty params}
\end{figure*}

\begin{figure*}[h]
    \centering
    \includegraphics[width=\textwidth]{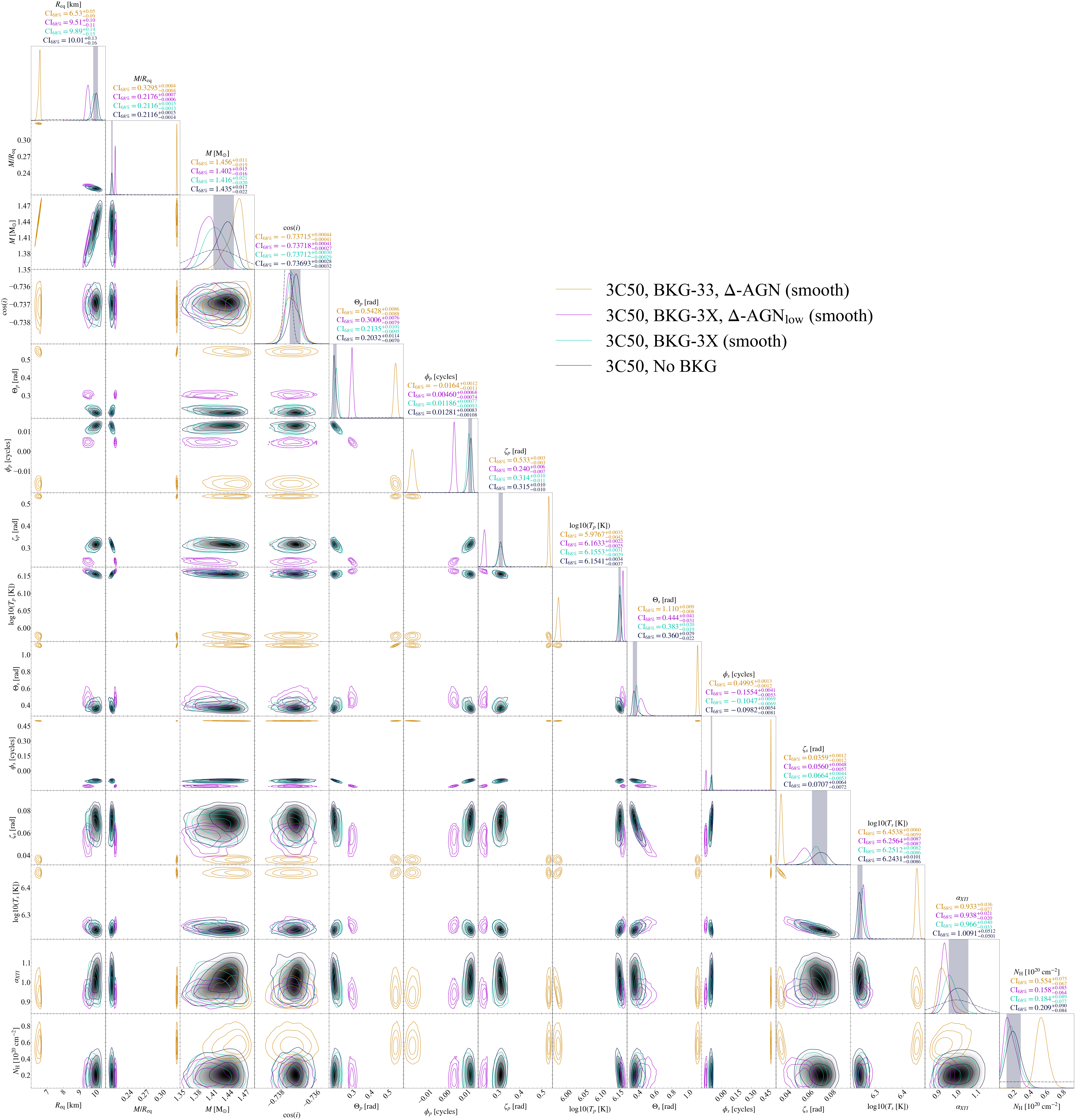}
    \caption{Posteriors obtained by \texttt{ST-U} as applied to the 3C50 dataset, testing the different implementations of background constraints.}
    \label{appendixfig: ST-U 3C50 bkg params}
\end{figure*}

\begin{figure*}[h]
    \centering
    \includegraphics[width=\textwidth]{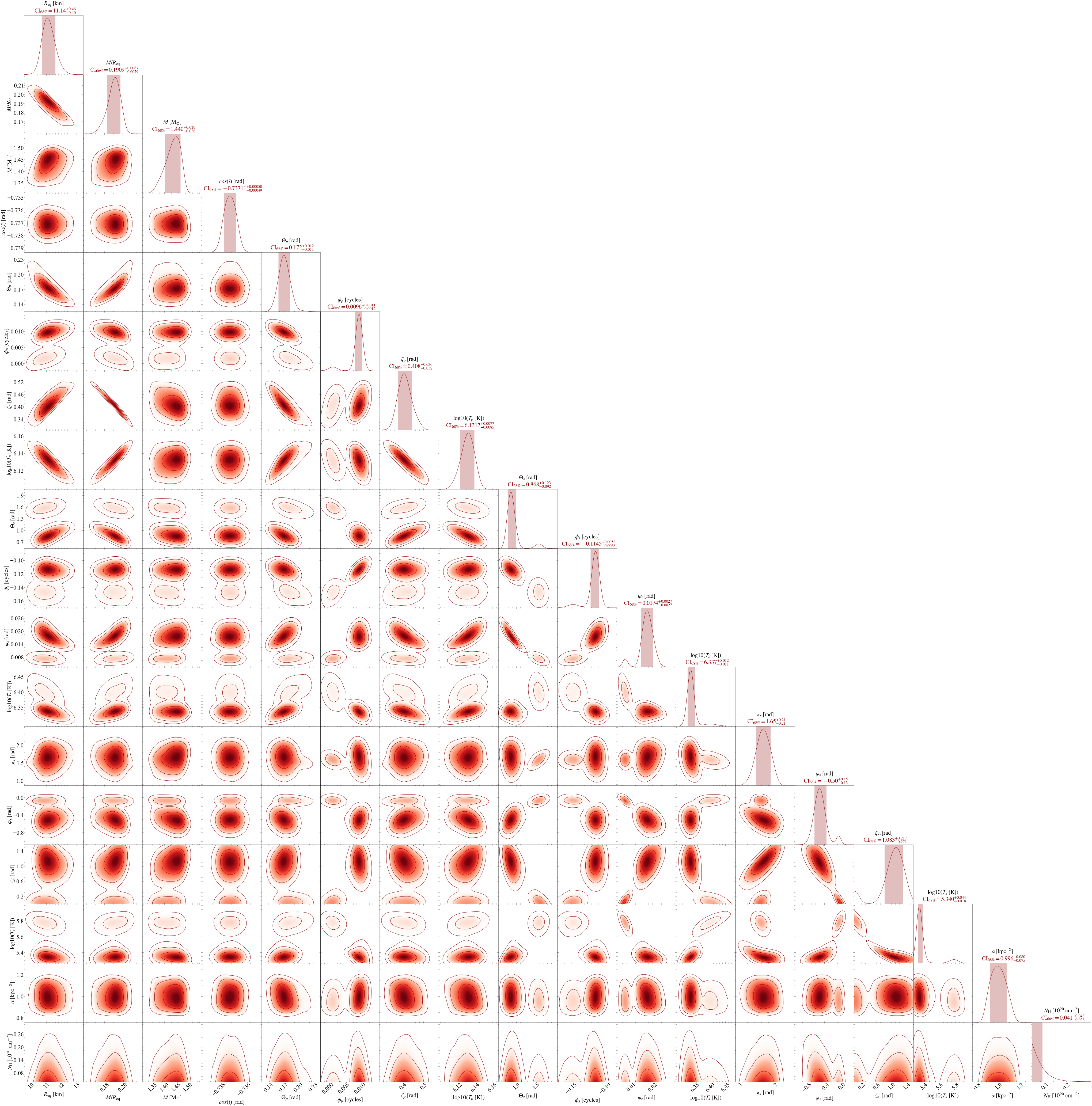}
    \caption{Posteriors obtained by \texttt{ST+PDT} as applied to the Delta dataset with lower background constraints imposed by the space weather estimate.}
    \label{appendixfig: ST+PDT Delta params}
\end{figure*}

\begin{figure*}[h]
    \centering
    \includegraphics[width=\textwidth]{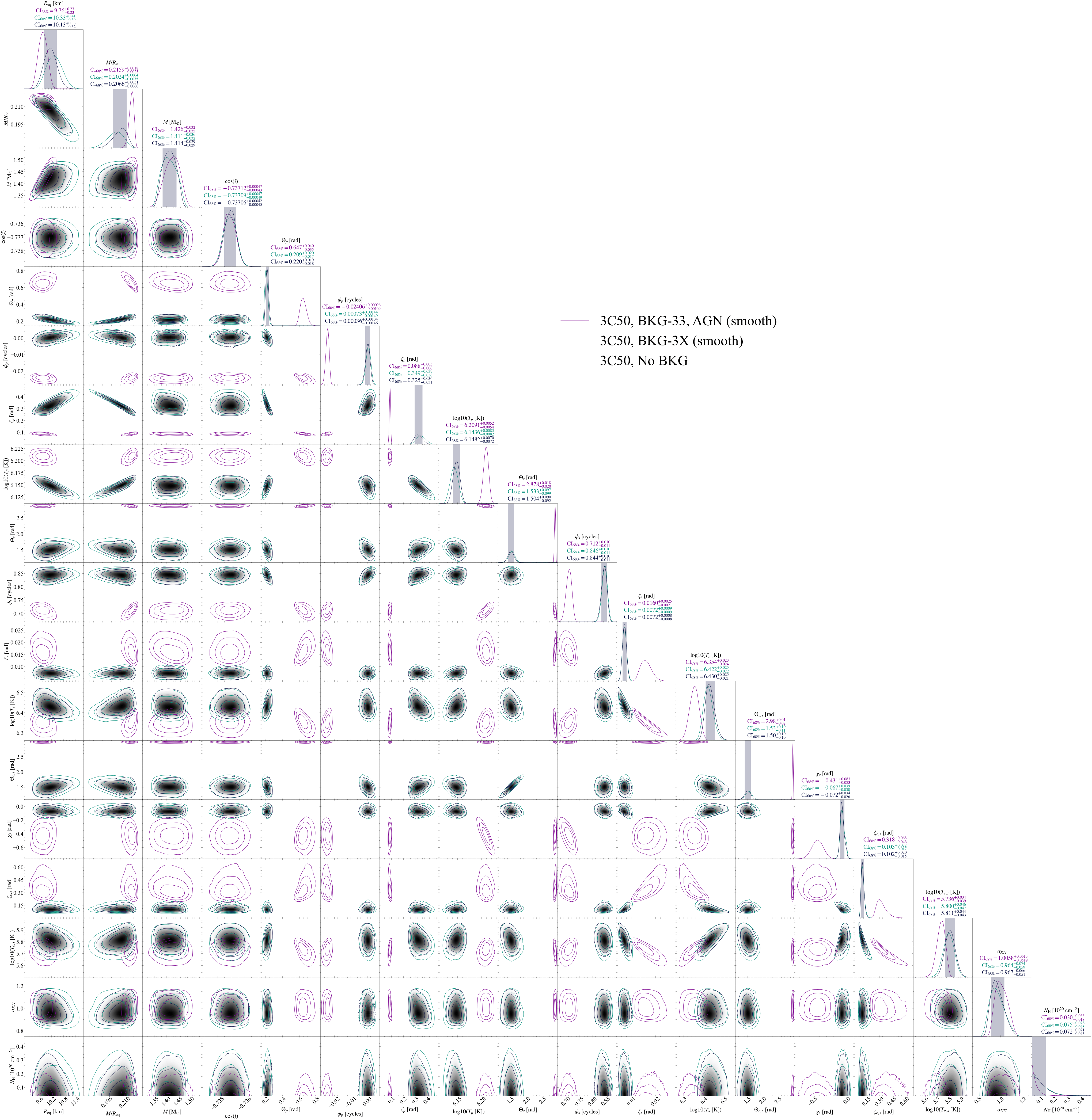}
    \caption{Posteriors obtained by \texttt{ST+PDT} as applied to the the 3C50 dataset, testing the different implementations of the 3C50 background constraints.}
    \label{appendixfig: ST+PDT 3C50 params}
\end{figure*}

\begin{figure*}[h]
    \centering
    \includegraphics[width=\textwidth]{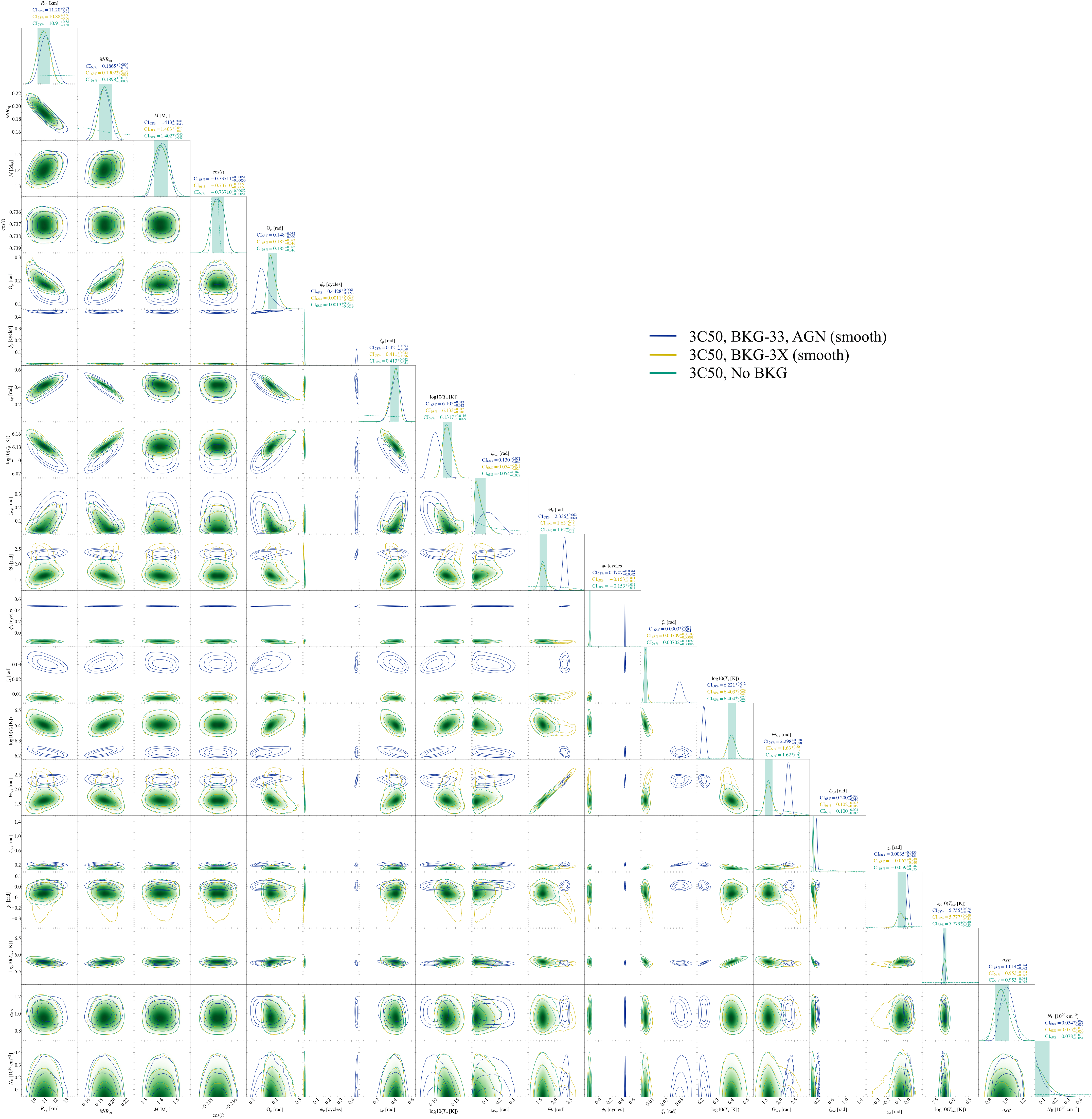}
    \caption{Posteriors obtained by \texttt{CST+PDT} (default resolution run) as applied to the the 3C50 dataset, testing the different implementations of the 3C50 background constraints.}
    \label{appendixfig: CST+PDT params}
\end{figure*}

\begin{figure*}[h]
    \centering
    \includegraphics[width=\textwidth]{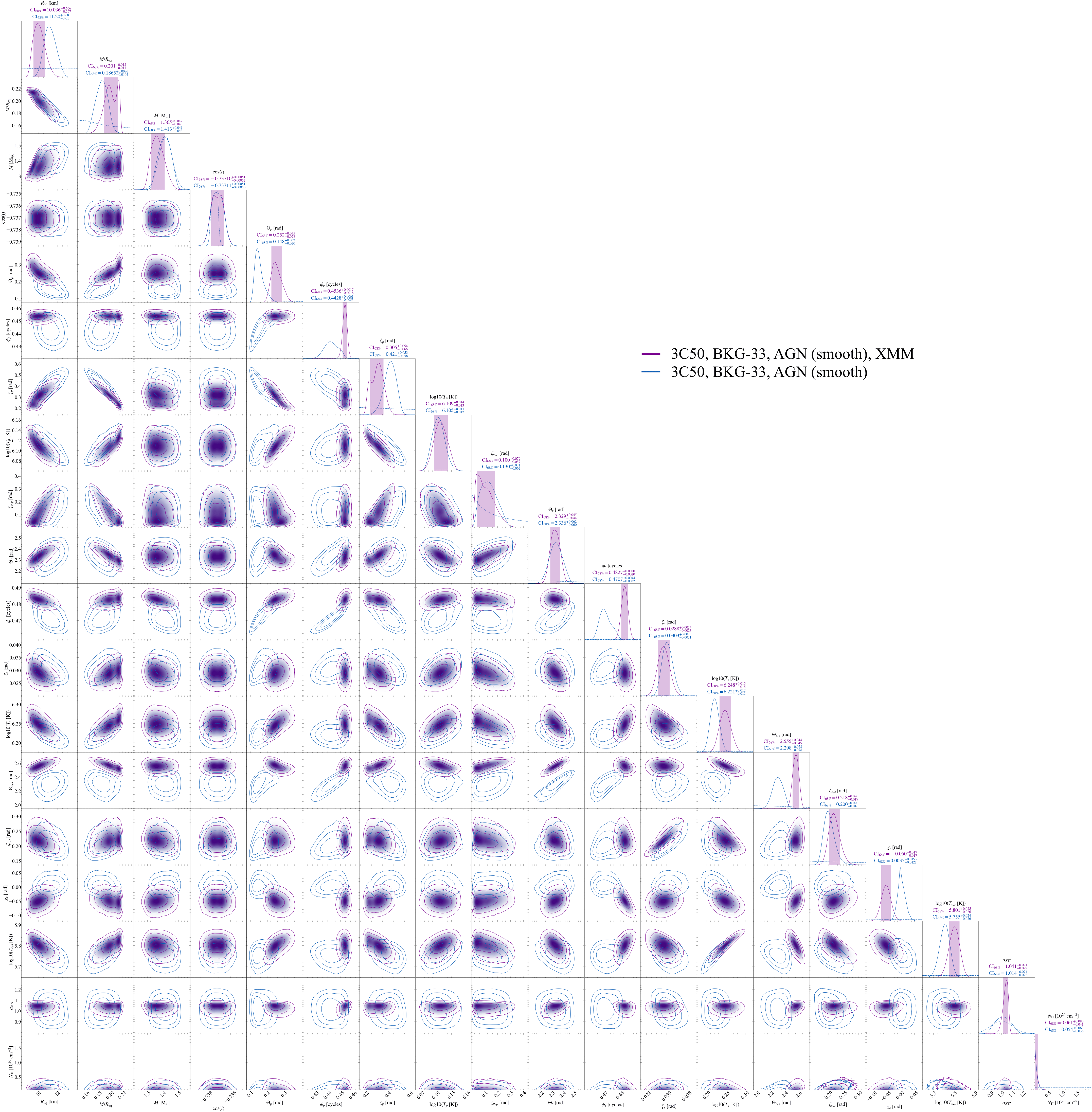}
    \caption{Posteriors obtained by \texttt{CST+PDT} for the NICER-only (default resolution run) involving both lower and upper background limits, and for the joint NICER-XMM run. }
    \label{appendixfig: CST+PDT XMM params}
\end{figure*}
\end{document}